 \documentclass{sig-alternate-2013-modified}
 \usepackage{latexsym}

\usepackage{amsmath}
\usepackage{graphicx}
\usepackage{times}
 \usepackage{ifpdf}
 \ifpdf\fi
 \usepackage{subfigure}
\usepackage{rotating}
\usepackage{multirow}
 \usepackage{url}

\usepackage{algorithm}
\usepackage{algorithmicx}
\usepackage[noend]{algpseudocode}

\usepackage{epsfig}
 \usepackage{times}

\newcommand{\ignore}[1]{}
\newcommand{\signore}[1]{}

\newcommand{\notinproc}[1]{#1}
\newcommand{\onlyinproc}[1]{}
\newcommand{\notinsub}[1]{}
\newcommand{\onlyinsub}[1]{#1}

\newtheorem{thm}{Theorem}[section]
\newtheorem{theorem}{Theorem}[section]

\newtheorem{lemma}[thm]{Lemma}

\def\deter{{\sc \phi}}  

\def\range{\mbox{{\sc rg}}}

\def\cS{{\cal S}}

\def\pr{\mbox{\sc pr}}
\def\var{\mbox{\sc var}}

\def\U{U*}
\def\L{L*}
\def\E{{\textsf E}}

\def\vecphi{\boldsymbol{\phi}}
\def\vecv{\boldsymbol{v}}
\def\vecu{\boldsymbol{u}}
\def\vecz{\boldsymbol{z}}

\newfont{\mycrnotice}{ptmr8t at 7pt}
\newfont{\myconfname}{ptmri8t at 7pt}
\let\crnotice\mycrnotice%
\let\confname\myconfname%

\permission{Permission to make digital or hard copies of all or part of this work for personal or classroom use is granted without fee provided that copies are not made or distributed for profit or commercial advantage and that copies bear this notice and the full citation on the first page. Copyrights for components of this work owned by others than the author(s) must be honored. Abstracting with credit is permitted. To copy otherwise, or republish, to post on servers or to redistribute to lists, requires prior specific permission and/or a fee. Request permissions from permissions@acm.org.}
\conferenceinfo{KDD'14,}{August 24--27, 2014, New York, NY, USA. \\
{\mycrnotice{Copyright is held by the owner/author(s). Publication rights licensed to ACM.}}}
\copyrightetc{ACM \the\acmcopyr}
\crdata{978-1-4503-2956-9/14/08\ ...\$15.00.\\
http://dx.doi.org/10.1145/2623330.2623680}

\begin{document}

\title{Distance Queries from Sampled Data: \\ Accurate and Efficient}

\numberofauthors{1}
\author{
\alignauthor Edith Cohen\\
\affaddr{Microsoft Research}\\
\affaddr{Mountain View, CA, USA}\\
       \email{editco@microsoft.com}
}

 \ignore{
\author{
Edith Cohen\thanks{Microsoft Research, Silicon Valley, USA}
  }
  }


 \maketitle
\begin{abstract}
  \small 
\notinproc{\footnote{This is a full version of a KDD 2014 paper.}}
Distance queries are a basic tool in data analysis.  They are used for
detection and localization of change for the purpose of anomaly
detection, monitoring, or planning.  Distance queries are particularly
useful when data sets such as measurements, snapshots of a system,
content, traffic matrices, and activity logs are collected repeatedly.

Random sampling, which can be efficiently performed over streamed or
distributed data, is an important tool for scalable data analysis. The
sample constitutes an extremely flexible summary, which naturally
supports domain queries and scalable estimation of statistics, which
can be specified after the sample is generated.  The effectiveness of
a sample as a summary, however, hinges on the estimators we have. 

We derive novel estimators for estimating $L_p$ distance from sampled
data.  Our estimators apply with the most common weighted sampling
schemes: Poisson Probability Proportional to Size (PPS) and its fixed
sample size variants.  They also apply when the samples of different
data sets are independent or coordinated.  Our estimators are
admissible (Pareto optimal in terms of variance) and have compelling properties.  

We study the performance of our Manhattan and Euclidean distance
($p=1,2$) estimators on diverse datasets, demonstrating  scalability
and accuracy even when a small fraction of the data is sampled. Our
work, for the first time, facilitates effective distance estimation over sampled data.

\end{abstract}




\section{Introduction} \label{intro:sec}

Data is commonly generated or collected repeatedly, where each {\em instance} has the form of a value
assignment to a set of {\em keys}:
Daily summaries of the number of queries containing certain keywords,
activity in a social network,
transmitted bytes  for IP flow keys,
performance parameters (delay, throughput, or loss)
for IP source destination pairs, environmental
measurements for sensor locations,  and requests for 
resources.  In these examples, 
each set of values (instance) corresponds to a particular time or
location.  The universe of possible key values is fixed across
instances but the values of a key are different.

  One of the most basic operations 
in data analysis are {\em Distance queries}, which are used to
detect, measure,  and localize  change \cite{Chawathe:SIGMOD1996,Kifer:VLDB2004}. 
  Their applications include
anomaly detection, monitoring, and planning. 
Formally, a difference query between instances is specified by
a meta-data based selection predicate that is applied to the keys.
The result is the distance between the vectors
projected on the selected keys.
The simple example in Figure~\ref{diff:example} shows two instances on
7 keys, and some queries specified on this data using Euclidean and Manhattan distances.


\begin{figure} [h]
 {\scriptsize 
\begin{center}
 \begin{minipage}{1.61in}
\begin{tabular}{|l|r|r|r|r|r|r|}
\multicolumn{6}{c}{{\bf data set}}\\
\hline 
key $h$: & $a$ & $b$ & $c$ & $d$ & $e$ & $f$ \\
\hline
 $v_1(h)$ & 5 & 0  & 4 &  5 & 8 & 7 \\
 $v_2(h)$ & 7 & 10 & 3 &  0 & 6 & 7 \\
\hline
\end{tabular}
 \end{minipage}
\end{center}
\ignore{
 \begin{minipage}{1.7in}
\begin{tabular}{|l|r|r|r|r|r|r|}
\hline 
\multicolumn{7}{|c|}{{\bf range function for key $h$:}}\\
\multicolumn{7}{|c|}{$\range_p(h)=|v_1(h)-v_2(h)|^p$}\\
\hline
key $h$: & $a$ & $b$ & $c$ & $d$ & $e$ & $f$ \\
\hline
$\range_1$ & 2 & 10 & 1 & 5 & 2 & 0 \\
$\range_2$ & 4 & 100 & 1 & 25 & 4 & 0 \\
\hline
\end{tabular}
 \end{minipage}
}
}

{\small
{\bf $L_p$-distance }  for $H \subset \{a,b,c,d,e,f\}$:
$L_{p}(H)=(L^p_{p}(H))^{\frac{1}{p}}$ \\
where $L^p_{p}(H)= \sum_{h\in H} |v_1(h)-v_2(h)|^p$ 
}

{\scriptsize {\bf Example Queries:}\\
\begin{tabular}{ll}
$L_1(\{a,\ldots,f\})  =  20$ & $L_{2}(\{a,\ldots,f\}) =  \sqrt{134}\approx 11.6$\\
$L_1(\{d,e,f\})  =  7$ & 
$L_2^2(\{a,e\})=4$  
\end{tabular}
}

\caption{Distances between two instances, 
Table shows a data set with two instances $i\in\{1,2\}$ and 7 keys
$\{a,\ldots,f\}$ and the 
values $v_i(h)$ of key $h$ in instance $i$.
The figure also provides example distance queries, specified for a
selected set $H$ of keys.
 \label{diff:example}}
\end{figure}

The collection and warehousing of massive data is subject to limitations on
storage, throughput, and bandwidth.
 Even when the data is stored in full, exact
processing of queries may be slow and resource consuming.
Random sampling of datasets is widely used as a
means to obtain a flexible summary over which we can query the data
while meeting these limitations
\cite{Knu69,Vit85,Broder:CPM00,BRODER:sequences97,ECohen6f,GT:spaa2001,Gibbons:vldb2001,BHRSG:sigmod07,DasuJMS:sigmod02,BHRSG:sigmod07,HYKS:VLDB2008,bottomk:VLDB2008,DLT:jacm07,varopt_full:CDKLT10,CK:sigmetrics09}.


The sampling scheme is applied to our data, and a set of random bits,
and returns a small subset of the entries.
 Our sample includes these entries and their values and has a fraction 
 of the size of the original dataset.  
The sample only includes nonzero entries, this is
particularly important when the data is sparse, meaning that the vast
majority of keys have  associated value $0$.
When the values distribution is skewed, we often apply
{\em weighted} sampling, 
meaning that the probability a key is sampled depends on its value  -- 
Favoring heavier keys allow for more
accurate estimates of sums and other statistics.  

Perhaps the most basic sampling scheme
is Poisson sampling, where keys are sampled independently. 
Weighting is often done using
Probability Proportional to Size (PPS)
sampling~\cite{Hajekbook1981}, where the inclusion probability is
proportional to the value.  
Other common sampling schemes are bottom-$k$ (order) samples, which
have the advantage over independent sampling of
yielding a sample size of exactly $k$.  Bottom-$k$ sampling generalizes
reservoir sampling and includes Priority (Sequential
Poisson) sampling and weighted sampling without replacement
~\cite{Rosen1997a,Rosen1972:successive,Ohlsson_SPS:1998,bottomk07:ds,DLT:jacm07,bottomk:VLDB2008,CK:sigmetrics09}.
These sampling schemes are very efficient to apply also when the 
data is streamed or distributed.

  Once we have the sample, we can quickly process approximate
  queries posed over the original data.  This is done by applying an
{\em estimator} to the sample. 
We seek estimators that provide good results when a small fraction of the data
is sampled.  In particular, we would like them to be {\em admissible},
that is, optimally use the information we have in
the sample, and be efficient to compute.   When estimating nonnegative
quantities, such as distances, we are interested in nonnegative estimators.

Consider the basic problem of estimating, for a given selection
predicate, the sum of values of keys selected by the predicate.
 This problem is solved well by
the classic Horvitz-Thompson (HT) \cite{HT52} estimator.
The estimate is the sum, over sampled
keys $i$ satisfying the predicate, of the 
ratio $v_i/p_i$, where $v_i$ is the value and $p_i$ is the 
 inclusion probability of $i$.  This {\em inverse-probability}
estimate is clearly unbiased (if $v_i>0
\implies p_i>0$): if the key is
not sampled, the estimate is $0$ and otherwise it is $v_i/p_i$,  giving
expectation $v_i$.  The estimate is also nonnegative when values are.
Moreover, the inverse-probability 
estimator is a UMVUE (uniform minimum variance unbiased estimator),
meaning that among all unbiased and nonnegative estimators, it 
minimizes variance point wise.  To apply this estimator, we need the
inclusion probability $p_i$ to be available to it when $v_i$ is.  This
is the case with both Poisson and bottom-$k$ sampling, when
implemented correctly.  

The HT estimate is a sum estimator:  conceptually, we apply an estimator to each key
in the universe.  Keys that are not sampled have an estimate of $0$.
Keys that are sampled have a positive estimate.  We then sum the
estimators of different keys.    This makes it highly suitable
for domain (selection) queries, which aggregate over a subset of keys
that can be specified by a predicate.

 We now turn to our problem of
estimating the distance between instances from their samples.
We seek an estimator with similar properties to the HT estimator:  a
sum estimator, unbiased and nonnegative, and with optimal use of the 
information in the sample.

This problem turns out to be significantly more challenging.  One can
attempt to apply again the HT estimator: When the outcome reveals the
value of the estimated quantity, the estimate is equal to the value
divided by the probability of such an outcome.  The estimate is $0$
otherwise.  Inverse probability estimates, however, are inapplicable
to distance estimation over weighted samples, since they require that
there is a positive probability for an outcome which reveals the exact value
of the estimated quantity:  the absolute difference between the values
of the key in two instances.  With multiple instances and weighted
sampling, keys that have zero value in one instance and positive value
in another have positive contribution to the distance but because zero
values are never sampled, there is zero probability for determining
the value from the outcome: In the Example in Figure
\ref{diff:example}, key $c$ has value $5$ in the first instance and
value $0$ in the second, and thus has a contribution of $5$ to the
L$_1$ distance.  The key however will never be included in a weighted
sample of instance 2.


When considering multiple 
  instances, we  also need to specify 
how their samples relate to each other.
  The sampling schemes we mentioned, Poisson and bottom-$k$, specify
  the distribution for a single instance.   Our first requirement, since
data of different instances can be distributed or collected in
 different times, is that the sample of one instance can  not depend on values assumed in
another~\cite{multiw:VLDB2009,CK:pods11}.  The random bits, however, can be
reused (using random hash functions), to make sampling probabilities dependent.
The two extremes of the joint
distribution of samples of different instances  are {\em independent sampling} (independent sets of random bits
for each instance) and {\em coordinated sampling} (identical sets of random
bits).  With coordinated sampling, which is a form of
locality sensitive hashing, similar instances have
similar samples whereas independent samples of identical instances
can be completely disjoint.  
Each of these two relations has unique advantages  and therefore in our
work here we address both:
Coordination
\cite{BrEaJo:1972,Saavedra:1995,Ohlsson:2000,Rosen1997a,Broder:CPM00,BRODER:sequences97,ECohen6f,GT:spaa2001,Gibbons:vldb2001,BHRSG:sigmod07,bottomk:VLDB2008,HYKS:VLDB2008,CK:sigmetrics09,multiw:VLDB2009}
allows for tighter estimates of many basic queries including distinct counts (set
union) \cite{ECohen6f,GT:spaa2001,Gibbons:vldb2001,CK:sigmetrics09},
quantile sums \cite{multiw:VLDB2009}, Jaccard similarity
\cite{Broder:CPM00,BRODER:sequences97}, and more recently L$_1$ distance
\cite{multiw:VLDB2009}.  The drawbacks of coordination
are that
it results in
unbalanced ``burden'' where same keys tend to be sampled across
instances -- an issue when, for example, being sampled translates to
overhead which would like to balance across keys.
Moreover, while beneficial for some
queries, the variance on other queries --
notably sum queries that span values from multiple instances  (``total
number of search queries by Californians on
  Monday-Wednesday'', from daily summaries) -- is
larger than with independent sampling -- an issue if our samples are
primarily used for such queries.

\medskip
\noindent
{\bf Contributions:}

We derive unbiased nonnegative estimators for $L^p_p$ distance
queries over weighted samples.  These include the two important cases of
Manhattan distances $L_1$ and Euclidean distances $L_2$ (which is the
square root of $L_2^2$).  
Our estimators apply with either 
Poisson or bottom-$k$ sampling.  We also address the two important cases where 
the samples of different instances are independent or coordinated. 
Our work facilitates, for the first time, the use of weighted samples 
as summaries that support distance queries. 

Our estimators have several compelling 
properties. 
Similarly to the HT estimator, our estimators are
 {\em sum} estimators:   We estimate $L^p_p$ as a sum, over
{\em the selected keys},  of nonnegative
unbiased estimates of $\range_p=|v_1-v_2|^p$  of the values assumed
by the key (see Figure\ref{diff:example}). 
Our estimators are unbiased, nonnegative, and admissible (Pareto optimal), 
meaning that another (nonnegative and unbiased) estimator with strictly 
lower variance on some data must have strictly higher variance on 
another data.  
The estimate $\hat{\range}_p$ obtained for a particular key has high
variance, since most likely, the key is not sampled in any instance
(in which case the estimate is $0$),
but unbiasedness allows for diminishing relative error when more keys
are selected.
The distance  $L_p$ can be
estimated by the $p$th root of our $L^p_p$ estimate.  

Our estimates $\hat{\range}_p$ can be positive also when the samples
do not reveal the respective value $\range_p$ but only partial
information on it.  This property turns out to be critical for
obtaining admissible estimators, what is not possible, as we mentioned
above, with the inverse-probability
estimate.  It also means, however,  that the estimators we derive
have to be carefully tailored to the power $p$.


 For independently-sampled instances, we present an
estimator for $\range_p$ of two values ($p>0$).  This derivation uses
a technique we presented in \cite{CK:pods11}.  Our estimator, which we
call the $L^*$ estimator, is the unique symmetric and {\em monotone}
admissible estimator, where monotonicity means
that the estimate is non-decreasing with the information we can glean from 
the  outcome.


 For coordinated samples of instances, we apply our framework of
monotone sampling~\cite{sorder:PODC2014}. We derive 
the \L\ and \U\ estimators for the $\range_p$ functions, which are
admissible, unbiased, and nonnegative.
 The \L\ estimator is monotone and has lower variance for 
data with small difference (range) whereas the \U\ estimator performs
better when the range is large. 
This choice is important,  because it allows us to customize
estimation to properties of the data set.  Network traffic data,
for example, is likely to have larger differences and thus the \U\
estimator may perform better whereas the \L\ estimator would be
preferable when differences are smaller.  The \L\ estimator also 
exhibits a compelling theoretical property of
being  ``variance competitive'' \cite{CKsharedseed:2012}, meaning that 
for all data vectors, its
variance is not too far off the minimum possible variance for the
vector by a nonnegative unbiased estimator.    This property makes the
\L\ estimator a good default choice.

For $p=1,2$,  which are the important special cases of the Manhattan
and the Euclidean distances, we compute closed form expressions of
estimators and their variance and also obtain tighter bounds on the
``competitiveness'' of the \L\ estimator.
We evaluate and compare the  performance of our $L_1$ and $L_2^2$ distance
estimators on queries over diverse data sets.  The queries vary in the support
size (number of keys in the data set satisfying the selection
predicate) and in the relative difference (difference normalized by norm).    We show that in
all cases, we achieve good results when a small fraction of the data is
sampled.  Over coordinated samples, 
we examine the behavior of the \L\ and \U\ estimators 
and also consider  the {\em optimally
  competitive} estimator, which minimizes the worst-case ratio, and we
compute by a program.
Finally, we provide guidelines
to choosing between these estimators based on properties of the data.



\medskip
\noindent
{\bf Roadmap:}  Section~\ref{prelim:sec} contains necessary
background and definitions.  We present difference estimators
for independent samples in Section~\ref{ind:sec} and
for coordinated samples in Section~\ref{exsharedests:sec}.
Section~\ref{dataeval:sec} contains an
experimental evaluation.

\section{Preliminaries} \label{prelim:sec}

We denote by $v_{ih} \in  \mathbb{R}_{\geq 0}$  the value of key $h\in K$ in instance
$i\in [r]$ and
by the vector $\vecv(h)$,  the values of key $h$ in all instances (the column vector
$v_{ih}$).
The {\em exponentiated range}  of a vector
$\vecv$ is:
\begin{equation} \label{exrangedef}
\range_p(\vecv)=\left(\max(\vecv)-\min(\vecv)\right)^p\, \,  \text{($p>0$) }
\end{equation}
where $\max(\vecv)\equiv \max_{i} v_i$ and $\min(\vecv)=\min_i v_i$ are the maximum and minimum entry values of the vector $\vecv$. We omit the subscript when $p=1$.

 We are interested in queries which specify a
selected subset $H\subset K$ of keys, through a predicate on $K$, and return
\begin{equation}
  L^p_p(H) = \sum_{h\in H} \range_p(\vecv(h))\ .
\end{equation}
The $L_p$-distance of two instances ($r=2$)
is $L_p(H)\equiv (L^p_p(H))^{1/p}$.

When data is sampled, we estimate 
$L^p_{p}$, by summing estimates $\hat{\range}_p$ for the respective single-key
primitive $\range_p(\vecv(h))$ 
over keys $h\in H$.  
We use nonnegative
unbiased estimators for the primitives, which result, from linearity of
expectation, in unbiased estimates for the sums. 
We measure error by the
coefficient of variation (CV), which 
is the ratio of the square root of the variance to the mean. 
Our estimates for each key have high variance, but
when inclusions of different keys are pairwise independent, variance
is additive and the CV decreases with $|H|$, allowing
for accurate estimates of the sum.
Finally, we can estimate $L_p(H)$
by taking the $p$th root of the estimate for
$L^p_p(H)$.  This estimate is biased,
but the error is small when the CV of
our $L^p_p(H)$ estimate is small.

 A basic component in applying sum estimators is obtaining from
 basic sampling schemes of instances the  respective estimation
 problems  for a {\em single key}.  
We cast the basic sampling schemes discussed in the introduction
in the following  form, which facilitates separate treatment of each key.

The sampling of the entry $v_{ih}$ is specified
by a threshold value
$\tau_{ih}\geq 0$, and random {\em seed} values
$u_{ih} \sim U[0,1]$ chosen uniformly at random. 
\begin{equation} \label{singlekey:eq}
\mbox{$h$ is sampled in instance $i$}\, \iff \,  v_{ih}\geq \tau_{ih} u_{ih}\ . 
\end{equation}

The estimation problem for a single key $h$ is then as follows.  We
know $\mbox{\boldmath{$\tau$}}=(\tau_{1h},\ldots,\tau_{rh})$,
the seeds $\vecu=(u_{1h},\ldots,u_{rh})$, and the results of the
sampling for $h$ (the value $v_{ih}$ in all instances $i$ where $h$
was sampled).  We apply an estimator $\hat{\range}_p$ to this
information, which we refer to as the {\em outcome} $S$,
 to estimate $\range_p(\vecv)$.  
The availability of the seeds $u_{ih}$ to the estimator turns out to be critical for estimation quality \cite{CK:pods11,sorder:PODC2014}.   We 
facilitate it by generating the seeds using random hash functions 
(pairwise independence between keys suffices for variance bounds).
When we treat a single key
$h$, we omit the reference to $h$ from the notation.

\medskip 
\noindent 
{\bf Sampling scheme of instances.}
We now briefly return to
 sampling schemes of instances, and show how we obtain the single-key
 formulation \eqref{singlekey:eq}  from them.

  With Poisson PPS (Probability Proportional to
Size) sampling of instance $i$, each key $h$ is sampled with
  probability proportional to  $v_{ih}$.  An equivalent formulation is
  to use a global threshold value $T_i$, such that a key $h$ is sampled if and only if
$v_{ih} \geq u_{ih} T_i$. The expected
sample size $\E[|S|]=\sum_{h\in K} \min\{1,v_{ih}/T_i\}$ is determined
by $T_i$.
The sampling can be
easily implemented with respect to either a desired sample size or a
fixed threshold value.
The respective estimation problem \eqref{singlekey:eq}  for 
key $h$ has  $\tau_{ih} \equiv T_i$.
As an example, we can obtain a PPS Poisson sample of expected size $\E[|S|]= 3$ for the instances in Figure~\ref{diff:example} 
using $T_1=29/3$ (instance 1) and $T_2=33/3=11$ (instance 2).

 Priority (sequential Poisson) sampling~\cite{Ohlsson_SPS:1998,DLT:jacm07,Szegedy:stoc06}
is performed by assigning each key a {\em priority}
$r_{ih}=v_{ih}/u_{ih}$. The sample of instance $i$ includes 
the $k$ keys with largest priorities, the $(k+1)^{\mbox{\small th}}$ largest priority
$T_i$, and the $k^{\mbox{\small th}}$ largest priority $T'_i$.

To obtain the single-key formulation \eqref{singlekey:eq} for key $h$,
we consider the sampling {\em conditioned} on fixing
the seeds $u_{ij}$ (and thus the priorities $r_{ij}$) for all
$j\not=h$~\cite{bottomk:VLDB2008,DLT:jacm07}.  
The {\em effective}  threshold $\tau_{ih}$ is
the $k^{\mbox{\small th}}$ largest priority in $K\setminus\{h\}$:
 Key $h$ is then sampled if and only if
$v_{ih}/u_{ih}> \tau_{ih}$.  When $h\in S$, $\tau_{ih}=T_i$ and when
$h\not\in S$, $\tau_{ih}=T'_i$.

When sampling several instances, we can make the samples
{\em independent} when we use independent
$u_{ih}$ for all $i$.  The samples are {\em coordinated}
(shared-seed) if the same seed is used for the same key in all
instances, that is, $\forall h\in K, \forall i\in [r],
u_{ih}=u_{1h}\equiv u_h$.

Figure~\ref{example1b:fig}  shows Poisson PPS and priority samples,
independent and coordinated, obtained 
for the two instances in Figure~\ref{diff:example} from a random seed
assignment $u_{ih}$.  The figure also shows the
threshold values $\tau_{ih}$, which are
 available  to the estimators $\hat{\range}_p(S)$.
As an example, the outcome for key $4$, (the outcome is the input to the
estimator), is as follows. When  instances are independently Poisson
sampled the outcome includes: $v_1=5$ (key 4 is
sampled only in instance $1$ so we know $v_{14}$ exactly), $u_1=0.15$, $u_2=0.36$ (seeds available
by applying hash functions to the key), and $\tau_1=29/3$,
$\tau_2=11$ (with Poisson PPS sampling, the same threshold applies to
all keys in each instance).
With coordinated priority sampling of instances the outcome includes
$v_1=5$, $u=0.15$, $\tau_1=13.8$ (since key 4 is sampled in instance
$1$) and $\tau_2=30.4$ (since key $4$ is not sampled in instance $2$).

 \begin{figure*} [t]
{\scriptsize 
\begin{minipage}{3.0in}
\begin{tabular}{|l|r|r|r|r|r|r|}
\hline 
 & $1$ & $2$ & $3$ & $4$ & $5$ & $6$ \\
\hline 
Instance $1$ & 5 & 0  & 4 &  5 & 8 & 7 \\
Instance $2$ & 7 & 10 & 3 &  0 & 6 & 7 \\
\hline 
\multicolumn{7}{c}{$\quad$} \\
\multicolumn{7}{c}{Independent sampling, PPS}\\
\hline 
seeds $u_1$ & 0.23 & 0.29  & 0.84 &  0.15 & 0.58 & 0.19 \\
seeds $u_2$ & 0.81 & 0.17  & 0.48 &  0.36 & 0.15 & 0.49 \\
\hline 
 $v_1/u_1$ & 21.7 & 0  & 4.8  & 33.3 & 13.8 & 36.8 \\
$v_2/u_2$ & 8.6 & 58.8 & 6.25 & 0 & 41.7 & 14.3 \\
\hline 
\multicolumn{7}{c}{$\quad$} \\
\multicolumn{7}{c}{Coordinated sampling, PPS} \\
\hline 
seeds $u$ & 0.23 & 0.29  & 0.84 &  0.15 & 0.58 & 0.19 \\
\hline 
 $v_1/u$ & 21.7 & 0 & 4.8 & 33.3 & 13.8 &  36.8 \\
 $v_2/u$ & 30.4 & 34.5 & 3.6 & 0 & 10.3 &  36.8 \\
\hline 
\end{tabular}
\end{minipage}
\begin{minipage}{1.5in}
Poisson samples $E[|S|]=3$:\\
\begin{tabular}{|l|l|l|}
\hline 
 & $\tau_i$ & $S$ \\
\hline 
\multicolumn{3}{c}{$\quad$}\\
\multicolumn{3}{c}{independent}\\
\hline 
1 &  $29/3$ & $\{1,4,5,6\}$ \\
2 & 11 & $\{2,5,6\}$ \\
\hline 
\multicolumn{3}{c}{$\quad$}\\
\multicolumn{3}{c}{coordinated $u\equiv u_1$}\\
\hline
1 & $29/3$ & $\{1,4,5,6\}$ \\
2 & 11   & $\{1,2,5,6\}$ \\
\hline
\end{tabular}
\end{minipage}
\begin{minipage}{1.1in}
priority samples $|S|=3$:\\
\begin{tabular}{|l|l|l|}
\hline
 & $\tau_{ih}$:  $h\in S$, $h\not\in S$  & $S$ \\
\hline
\multicolumn{3}{c}{$\quad$}\\
\multicolumn{3}{c}{independent}\\
\hline
1 & 13.8,\, \,  21.7 & $\{4,5,6\}$ \\
2 & 8.6,\, \,  14.3 & $\{2,5,6\}$ \\
\hline
\multicolumn{3}{c}{$\quad$}\\
\multicolumn{3}{c}{coordinated $u\equiv u_1$}\\
\hline
1 & 13.8,\, \,  21.7 & $\{4,5,6\}$ \\
2 & 10.3,\, \,  30.4 & $\{1,2,6\}$ \\
\hline
\end{tabular}
\end{minipage}
}
\caption{Independent and coordinated samples of two instances.  
Poisson PPS samples of expected size 3 and
  priority samples of size 3 ($k=3$).\label{example1b:fig}}
\end{figure*}

\ignore{
\medskip
\noindent
{\bf Sampling model (single key):} 
 The exponentiated range estimators are applied to samples of the same
 key $h$ across instances $i\in r$.   That is, we work with the
restriction of the sample to one key at a time.

With Poisson sampling, for key $h$, we can obtain
from the sample
\eqref{singlekey:eq},  the values of sampled entries of
key $h$.  The seed vector $\vecu\equiv \vecu(h)$ and the thresholds 
$\mbox{\boldmath{$\tau$}}=(\tau_1(h),\ldots,\tau_r(h))$ are all
available to the estimator.
With Bottom-$k$ sampling of instances, the threshold is not readily
available,
so we work with {\em effective} thresholds as follows.
We condition the inclusion of $h$ on seeds 
of other keys being fixed~\cite{bottomk:VLDB2008,DLT:jacm07} and define
$\tau^h_i \equiv \tau_i$ to be the inverse of the
$k$th largest $r_i(h)$  of
keys in instance $i$ with $h$ excluded (which is the $k+1$st largest
ratio over all keys in the instance).
}
  From here onward, we focus on
estimating $\range_p(\vecv)$ for a single key from the respective
outcome $S$. We
return to sum aggregates only for the experiments
in Section~\ref{dataeval:sec}.


\ignore{
Sampling is {\em independent} if $\tau_1,\ldots,\tau_r$ are  
independent, equivalently, if $u_1,\ldots,u_r$ are independent.
Sampling is {\em coordinated (shared-seed)} if $u_1=u_2=\cdots=u_r
\equiv u$.    Sampling is PPS
(proportional to size) if for some fixed vector
$\mbox{\boldmath{$\tau^*$}}$, $\tau_i = u_i \tau^*$ -- in which case
entry $i$ is included with probability
$\min\{1,v_i/\tau^*_i\}$. We refer to $\vecu$ as the {\em seed} vector.
The outcome $S\equiv S(\vecu,\vecv)$ is a function of the random seed(s)
and the data.

We assume that $\vecu\in [0,1]^r$, its distribution, and
the function $\mbox{\boldmath{$\tau$}}(\vecu)$ are available with the sample.
Seeds  are generated using random hash functions:
Applied to the key with coordinated sampling and to the key and
instance pair with independent sampling.
}

\smallskip
\noindent
{\bf Estimators:}
For an outcome $S$, we denote by
\begin{eqnarray}
S^* & = & \{\vecv\, \mid \, S=S(\vecu,\vecv)\} \nonumber\\
&=& \{\vecz \mid  i\in S \implies z_i=v_i \, , 
\,  i\not\in S \implies z_i< \tau_i u_i \}\  \label{sstardef}
\end{eqnarray}
the set of all data vectors consistent with $S$.
We can equivalently define the outcome as the set $S^*$
since it captures 
all the information available to the estimator on $\vecv$ and hence on 
$\range_p(\vecv)$.  For our example in Figure \ref{example1b:fig},
considering independent Poisson sampling and key $4$, the set
$S^*$ includes all vectors $(5,x)$ such that $x< u_2\tau_2=0.36\cdot 11 = 3.96$. The actual data vector for key $4$ is
$(5,0)$, but the outcome only partially reveals it.

We denote by $\mathcal{S}$ the set of all possible outcomes, that is,
any outcome consistent with any data
vector $\vecv$ in our domain.  For data $\vecv$, we denote by
$\mathcal{S}_{\vecv}$ the probability distribution over outcomes
consistent with $\vecv$.
As mentioned, we are seeking 
{\em nonnegative} estimators   $\hat{\range}_p(S)\geq 0$ for all $S\in
\cS$, since $\range_p$ are nonnegative.   
Since we sum many estimates, we would like each estimate to be
{\em unbiased} $\E_{S\sim S_{\vecv}}[\hat{\range}_p(S)]=\range_p(\vecv)$.  
We also seek 
{\em bounded variance} on all data $\vecv$, $\E_{S\sim
  S_{\vecv}}[\hat{\range}_p(S)^2]<\infty$,  and {\em admissibility}
(Pareto variance optimality): there is no nonnegative unbiased 
estimator with same or lower variance on all data
and strictly lower on some data.
An intuitive property that is sometimes desirable is {\em monotonicity}:
the estimate value is non decreasing with the
information on the data that we can glean from the outcome
$S^*\subset S'^*\, \implies\, \hat{\range}_p(S) \geq \hat{\range}_p(S')$.


When $S^*$ includes vectors $\vecv$ such that $\range_p(\vecv)=0$, any
unbiased and nonnegative estimator 
must have $\hat{\range}_p(S)=0$ (with probability $1$).   We therefore
limit our attention to estimators satisfying this property.

 We can also see that when
the key $h$ is not sampled in any instance, then $S^*$ is consistent
with $\range_p(\vecv)=0$, which means that $\hat{\range}_p(S)=0$ and
we therefore do not  need to explicitly compute the contribution 
of the vast majority of keys that are not sampled in at least one 
instance. 

 Finally, we will use the following definition of order optimality of
 estimators in our constructions.
Given a partial order $\prec$ on the data domain 
an estimator $\hat{f}$ is  $\prec$-optimal (respectively,
$\prec^+$-optimal) if it is unbiased (resp., and nonnegative)
for all data $\vecv$, and minimizes variance
for $\vecv$ conditioned on the variance being
minimized for all preceding vectors.
Formally, if there is no
other  unbiased (resp., nonnegative) estimator that
has strictly lower variance on some data $\vecv$ and at most the
variance of  $\hat{f}$ on all vectors that precede $\vecv$.
An order optimal estimator is admissible.



\section{Independent PPS sampling}  \label{ind:sec}

  We derive estimators for $\range_p(\vecv)$, where $\vecv=(v_1,v_2)$
(two instances $r=2$).  The estimation scheme is specified by
$\mbox{\boldmath{$\tau$}}=(\tau_1,\tau_2)$.  The
outcome $S(\vecu,\vecv)$ is determined by the data vector $\vecv=(v_1,v_2)$
and $\vecu=(u_1,u_2)$, where $u_i\sim U[0,1]$ are independent.
The set $S^*$ of vectors consistent with $S$ is \eqref{sstardef}.

  We derive the \L\ estimator, $\hat{\range}_p^{(L)}$, which is the
  unique symmetric, monotone,  and admissible estimator.  The construction
adapts a framework from \cite{CK:pods11}, which was
  used to estimate $\max\{v_1,v_2\}$:
  We specify an order $\prec$ on the data domain.  We then formulate a set of
sufficient constraints  for an unbiased symmetric and order-optimal estimator
$\hat{f}^{(\prec)}$ of $\range_p(\vecv)$.  The constraints, however,
do not incorporate nonnegativity, as this results in much more complex
dependencies.  But if we find a nonnegative
solution $\hat{f}^{(\prec)}$, then we find an estimator with all the desired properties.
  We therefore hope for a good ``guess'' of $\prec$.

We work with $\prec$ that prioritizes smaller distances, that is, 
$\vecv \prec \vecz$ if and only if 
  $\range(\vecv) < \range(\vecz)$.  
With each outcome $S\in \cS$, we associate its 
{\em determining vector} $\deter(S)$, which we define as the 
$\prec$-minimal vector in (the closure of) $S^*$.   The closure is the 
set obtained when using a non-strict inequality in \eqref{sstardef}.
The determining vector is unique for all outcomes $S$ that are not
consistent with $\range_p(\vecv)=0$, that is, 
$\range_p(\vecv)>0$ for all $\vecv\in S^*$.  As we mentioned earlier, 
we only need to specify the estimator on these outcomes, since we only
consider estimators that are $0$ on outcomes consistent with
$\range_p(\vecv)=0$. 
The mapping of outcomes $S$ to the determining vector $\deter(S)$ is shown in Table~\ref{rangedindest:table} (Bottom). 

 We now formulate sufficient constraints for  $\prec$-optimality.
Conditioned on fixing the
 estimator on outcomes $S$ such that $\deter(S) \prec \vecv$,  the
 ``best'' we can do, in terms of minimizing variance, is to set it
to a fixed
value on all outcomes such that $\deter(S)=\vecv$.  This fixed value
is determined by the unbiasedness requirement.
Since $\hat{f}^{(\prec)}$ is the same for all outcomes with 
  same determining vector,  we specify it as a function of 
the determining vector $\hat{f}^{(\prec)}(S) \equiv 
\hat{f}^{(\prec)}(\deter(S))$. 


We use the notation $\mathcal{S}_0(\vecv)$  for the set of  outcomes $S$ 
that are
consistent with $\vecv$ but also consistent with a vector that precedes $\vecv$:
\begin{eqnarray*}
\mathcal{S}_0(\vecv) & = & \{ S |\vecv\in S^* \, \wedge \, \deter(S) \prec\vecv\}
\end{eqnarray*}
 The contribution of the outcomes $\mathcal{S}_0(\vecv)$ to the
 expectation of $\hat{f}^{(\prec)}$ when data is $\vecv$  is
$$f_0(\vecv)=\E_{S\sim \mathcal{S}_{\vecv}}[I_{S \in \mathcal{S}_0(\vecv)} \hat{f}^{(\prec)}(S)]\
,$$
where $I$ is the indicator function.
 We obtain the following sufficient constraints
for a $\prec$-optimal unbiased estimator, that may not be
 nonnegative, but is forced to be $0$ on all outcomes consistent with $\range_p(\vecv)=0$.
 For all $\vecv$,
{\small
\begin{eqnarray} 
\pr_{S\sim \mathcal{S}_{\vecv}}[\deter(S)=\vecv]=0  &\implies& f_0(\vecv)\equiv
f(\vecv) \label{cond1ub} \\
\pr_{S\sim \mathcal{S}_{\vecv}}[\deter(S)=\vecv]>0  &\implies&
\hat{f}^{(\prec)}(\vecv) = \frac{f(\vecv)-f_0(\vecv)}{\pr_{S\sim {\cal
      S}_{\vecv}}[\deter(S)=\vecv]} \label{dominantest:eq}\ .
 \end{eqnarray}
}


\begin{table}[htbp]
{\small
\begin{tabular}{ll}
\hline
$\vecphi=(\phi_1,\phi_2)$ \, & $\hat{\range}_p^{(L)}(\vecphi)$ \\
\hline
$\vecphi=(0,0)$ \, & $0$ \\
$\phi_1 \geq \phi_2 > \tau_2$ \, &  $\frac{\tau_1}{\min\{\tau_1,\phi_1\}}(\phi_1-\phi_2)^p$ \\
$\phi_1 \geq \phi_2 \leq \tau_2$ \, & $\frac{p \tau_1
  \tau_2}{\min\{\phi_1,\tau_1\}}
\int_{\max\{0,\phi_1-\tau_2\}}^{\phi_1-\phi_2} \frac{y^{p-1}}{\phi_1-y}dy +$ \\
& $+\frac{\tau_1 \max\{0,\phi_1-\tau_2\}^p}{\min\{\phi_1,\tau_1\}}$
\end{tabular}
$\quad\quad$
\begin{tabular}{lcll}
\hline
outcome $S$ & & $\deter(S)_1$ & $\deter(S)_2$ \\
\hline
$S=\emptyset$ & : &  $0$ & $0$  \\
$S=\{1\}$ & : & $v_1$ & $\min\{u_2 \tau_2,v_1\}$    \\
$S=\{2\}$ & : & $\min\{u_1 \tau_1,v_2\}$  & $v_2$  \\
$S=\{1,2\}$ & : & $v_1$ & $v_2$
\end{tabular}

}
\caption{Top: Estimator $\hat{\range}_p^{(L)}$ for $p>0$ over
  independent samples, stated as a function of the determining vector $\vecphi=(\phi_1,\phi_2)$ when $\phi_1\geq \phi_2$ (case $\phi_2>\phi_1$ is symmetric). Bottom:
mapping of outcomes to determining vectors. \label{rangedindest:table}}
\end{table}

We derive $\hat{\range}_p^{(L)}$ by solving 
the right hand side of \eqref{dominantest:eq} for all $\vecv$ such that $\pr_{S\sim
  \mathcal{S}_{\vecv}}[\deter(S)=\vecv]>0$.
 The solution $\hat{\range}_p^{(L)}(\vecphi)$  ($p>0$) 
is provided in
Table~\ref{rangedindest:table} through a mapping of determining vectors to
estimate values. The estimator is specified for $\phi_1 \geq \phi_2$,
as the other case is symmetric.   We can verify that for all $p>0$, 
the estimator  $\hat{\range}_p^{(L)}$ is nonnegative, monotone
(for all $y$, $\hat{\range}_p^{(L)}(y,x)$ is non-increasing for $x\in(0,y]$) and
has finite variances (follow from $\int_0^y \hat{\range}_p^{(L)}(y,x)^2 dx < \infty$).
We can also verify that condition \eqref{cond1ub} holds.
Vectors $\vecv$ with 
$\pr_{S\sim 
  \mathcal{S}_{\vecv}}[\deter(S)=\vecv]=0$
are exactly those with one 
positive and one zero entry.  We can verify that \eqref{cond1ub} is 
satisfied, that is, 
$\E_{S \sim 
  \mathcal{S}_{\vecv}} \hat{\range}_p^{(L)}(S)=\range_p(\vecv)$
on these vectors.  
Table~\ref{rangeind2est:table} shows
explicit expressions of $\hat{\range}^{(L)}$ and ${\hat{\range}_2}^{(L)}$.


\begin{table}[htbp]
{\scriptsize
\begin{minipage}{3.3in}
\begin{tabular}{|l|l}
\hline
$\vecphi=(\phi_1,\phi_2)$  & $\hat{\range}^{(L)}(\vecphi)$ \\
\hline
$\vecphi=(0,0)$ & $0$ \\
$\phi_1 \geq \phi_2 > \tau_2$ &  $\frac{\tau_1}{\min\{\tau_1,\phi_1\}}(\phi_1-\phi_2)$ \\
$\phi_1 \geq \phi_2 \leq \tau_2$ &  $\frac{\tau_1 \tau_2}{\min\{\tau_1,\phi_1\}} \ln\bigg(\frac{\min\{\phi_1,\tau_2\}}{\phi_2}\bigg)+\frac{\tau_1 \max\{0,\phi_1-\tau_2\}}{\min\{\phi_1,\tau_1\}}$ 
\end{tabular}
\end{minipage}
\begin{minipage}{3.3in}
\begin{tabular}{|l|l}
\hline
$\vecphi=(\phi_1,\phi_2)$ &  $\hat{\range}_2^{(L)}(\vecphi)$ \\
\hline
$\vecphi=(0,0)$ & $0$ \\
$\phi_1 \geq \phi_2 > \tau_2$ &  $\frac{\tau_1}{\min\{\tau_1,\phi_1\}}(\phi_1-\phi_2)^2$ \\
$\phi_1 \geq \phi_2 \leq \tau_2$ &  $\frac{2\tau_1
  \tau_2}{\min\{\tau_1,\phi_1\}}\bigg(\phi_2-\min\{\phi_1,\tau_2\}+\phi_1\ln\frac{\min\{\phi_1,\tau_2\}}{\phi_2}
\bigg)$\\
& $+\frac{\tau_1\max\{0,\phi_1-\tau_2\}^2}{\min\{\phi_1,\tau_1\}}$
\end{tabular}
\end{minipage}
}
 \caption{Explicit form of estimators $\hat{\range}^{(L)}$ and
   $\hat{\range}_2^{(L)}$  for $r=2$ over independent samples.
   Estimator is stated as a function of the determining vector
   $(\phi_1,\phi_2)$ when $\phi_1\geq \phi_2$ (case $\phi_2\geq \phi_1$ is symmetric).\label{rangeind2est:table}}
 \end{table}

We now provide the derivation.
 We consider vectors $\vecv$
in increasing $\prec$ order and solve \eqref{dominantest:eq}  for $f^{(\prec)}$ on outcomes with
determining vector $\vecv=(v,v-\Delta)$, where $v\geq \Delta \geq 0$.


\begin{trivlist}
\item
$\bullet$ {\bf Case:}  $v-\Delta \geq \tau_2$.  The outcomes always
reveals the second entry.  The determining vector is $\vecv$ when
$u_1\tau_1 \leq v$, which happens with probability
$\min\{1,v/\tau_1\}$.
Otherwise, the outcome is consistent with $(v-\Delta,v-\Delta)$ and
the estimate is $0$.
We solve
the equality $\Delta^p = \min\{1,v/\tau_1\}\hat{\range}_p^{(L)}$, obtaining
\begin{equation}
\hat{\range}_p^{(L)}(v,v-\Delta)=\frac{\tau_1}{\min\{v,\tau_1\}}\Delta^p\ .
\end{equation}
\item
\ignore{
$\bullet$ {\bf Case:}  $v-\Delta < \tau_2$, $v \geq \tau_1$.
We have the equation
$$\Delta = \frac{v-\Delta}{\tau_2}\hat{\range}^{L}(v,v-\Delta)+\frac{1}{\tau_2}\int_{\max\{0,v-\tau_2\}}^{\Delta}\hat{\range}^{L}(v,v-y) dy\ .$$
We solve by using $g(\Delta)=\hat{\range}^{L}(v,v-\Delta)$, taking a partial derivative with respect to $\Delta$, solving for the derivative, and normalizing to find the constant.
We obtain
\begin{equation}
\hat{\range}^{L}(v,v-\Delta)=\tau_2 \ln\bigg(\frac{\min\{v,\tau_2\}}{v-\Delta}\bigg)
\end{equation}
\item
$\bullet$ {\bf Case:}  $v-\Delta < \tau_2$, $v \leq \tau_1$.
From unbiasedness,
$$\Delta=\frac{v}{\tau_1}\frac{v-\Delta}{\tau_2}\hat{\range}^{(L)}(v,v-\Delta) + \frac{v}{\tau_1\tau_2}\int_{\max\{0,v-\tau_2}^{\Delta}\hat{\range}^{(L)}(v,v-y)dy$$
Solving we obtain
\begin{equation}
\hat{\range}^{L}(v,v-\Delta)=\frac{\tau_1\tau_2}{v} \ln\bigg(\frac{\min\{v,\tau_2\}}{v-\Delta}\bigg)
\end{equation}
}
$\bullet$ {\bf Case:}  $v-\Delta < \tau_2$.
The determining vector is $(v,v-\Delta)$ with probability
$\frac{\min\{v,\tau_1\}}{\tau_1}\frac{v-\Delta}{\tau_2}$.  Otherwise,
it is $(v,v-y)$ for some $y<\Delta$.  We use
\eqref{dominantest:eq} to obtain an integral equation:
{\small
\begin{eqnarray*}
\Delta^p &=& \frac{\min\{v,\tau_1\}}{\tau_1}\frac{v-\Delta}{\tau_2}{\hat{\range}_p}^{(L)}(v,v-\Delta)+  \\
 && + 
\frac{\min\{v,\tau_1\}}{\tau_1\tau_2}\int_{\max\{0,v-\tau_2\}}^{\Delta}\hat{\range}_p^{(L)}(v,v-y)dy
\end{eqnarray*}
}
 Taking a partial derivative with respect to $\Delta$, we obtain
$$\frac{\partial \hat{\range}_p^{(L)}(v,v-\Delta)}{\partial \Delta} = \frac{p \tau_1 \tau_2}{\min\{v,\tau_1\}} \frac{\Delta^{p-1}}{v-\Delta}$$

 We use the boundary value for $\Delta=\max\{0,v-\tau_2\}$:
$$\hat{\range}_p^{(L)}(v,\min\{v,\tau_2\})=\frac{\tau_1}{\min\{v,\tau_1\}}
\max\{0,v-\tau_2\}^p\ ,$$ and obtain the solution
\begin{eqnarray}
 \lefteqn{\hat{\range}_p^{(L)}(v,v-\Delta)=}\\
 && 
\frac{p \tau_1 \tau_2}{\min\{v,\tau_1\}} \int_{\max\{0,v-\tau_2\}}^\Delta \frac{y^{p-1}}{v-y}dy + \frac{\tau_1 \max\{0,v-\tau_2\}^p}{\min\{v,\tau_1\}}\nonumber  \end{eqnarray}
\end{trivlist}


\noindent
{\bf The special case $\tau_1=\tau_2=\tau$:}
The estimators $\hat{\range}^{(L)}$ and $\hat{\range}_2^{(L)}$ as a
function of the determining vector and their variance are provided in
Tables~\ref{indL1L:table} and \ref{indL2L:table}.
\notinproc{
For data vectors where $v_1 \geq v_2 \geq \tau$, $\hat{\range}^{(L)}=v_1-v_2$
and $\var[\hat{\range}^{(L)}]=0$.
If  $v_1 \geq \tau \geq v_2$, $\hat{\range}^{(L)}=\tau\ln\frac{\tau}{v_2}+v_1-\tau$, and $\var[\hat{\range}^{(L)}]= -2\tau v_2\ln(\frac{\tau}{v_2})-v^2_2+(\tau)^2$.
 Finally, if
$v_2 \leq v_1 \leq \tau$,
$\hat{\range}^{(L)}(v_1,v_2)=\frac{(\tau)^2}{v_1}\ln\frac{v_1}{v_2}$ and
\begin{eqnarray*}
\lefteqn{\var_{\mathcal{S}_{\vecv}}[\hat{\range}^{(L)} ] =}\\
&=& \frac{v_1 v_2}{(\tau)^2}(\frac{(\tau)^2}{v_1}\ln \frac{v_1}{v_2} - (v_1-v_2))^2 +  \\
 & &
+ (1-\frac{v_1^2}{(\tau)^2})(v_1-v_2)^2 +  \\
 &&+ 
\frac{v_1}{(\tau)^2}\int_{v_2}^{v_1}(\frac{(\tau)^2}{v_1}\ln\frac{v_1}{y}-(v_1-v_2))^2 dy \\
&=& 2(\tau)^2(1-\frac{v_2}{v_1}\ln\frac{v_1}{v_2}-\frac{v_2}{v_1})-(v_1-v_2)^2\ .
\end{eqnarray*}
}
\begin{table} [th]
{\scriptsize
\begin{minipage}{3in}
\begin{tabular}{l|l}
\hline
Determining vector $\phi_1\geq \phi_2$ & $\hat{\range}^{(L)}(\vecphi)$ \\
\hline
$\phi_1 \geq \phi_2 \geq \tau$ & $\phi_1-\phi_2$\\
$\phi_1 \geq \tau \geq \phi_2$ & $\tau\ln\frac{\tau}{\phi_2}+\phi_1-\tau$\\
$\phi_2 \leq \phi_1 \leq \tau$ & $\frac{(\tau)^2}{\phi_1}\ln\frac{\phi_1}{\phi_2}$
\end{tabular}
\end{minipage}
\begin{minipage}{3in}
\begin{tabular}{l|l}
\hline
Data  $v_1\geq v_2$ & $\var_{\mathcal{S}_{\vecv}}[\hat{\range}^{(L)}]$ \\
\hline
$v_1 \geq v_2 \geq \tau$ & $0$ \\
$v_1 \geq \tau \geq v_2$ & $-2\tau v_2\ln(\frac{\tau}{v_2})-v^2_2+\tau^2$\\
$v_2 \leq v_1 \leq \tau$ & $2\tau^2(1-\frac{v_2}{v_1}\ln\frac{v_1}{v_2}-\frac{v_2}{v_1})-(v_1-v_2)^2$
\end{tabular}
\end{minipage}
}
\caption{$\hat{\range}^{(L)}$ and its variance for independent samples.\label{indL1L:table}}
\end{table}

\notinproc{
Similarly, for $v_2\leq v_1\leq \tau$,
${\hat{\range}_2}^{(L)}(v_1,v_2)=2(\tau)^2(
\ln\frac{v_1}{v_2}-\frac{v_1-v_2}{v_1})$.  }
\begin{table}
{\scriptsize
\begin{tabular}{l|l}
\hline
condition & $\hat{\range}_2^{(L)}(\vecphi)$ \\
\hline
$\phi_1 \geq \phi_2 \geq \tau$ & $(\phi_1-\phi_2)^2$\\
$\phi_1 \geq \tau \geq \phi_2$ & $\phi_1^2-(\tau)^2  -2\tau(\phi_1-\phi_2) +2\tau \phi_1\ln\frac{\tau}{\phi_2}$\\
$\phi_2 < \phi_1 \leq \tau$ & $2(\tau)^2( \ln\frac{\phi_1}{\phi_2}-\frac{\phi_1-\phi_2}{\phi_1})$
\end{tabular}
\begin{tabular}{l|l}
\hline
Data  $v_1\geq v_2$ & $\var_{\mathcal{S}_{\vecv}}[\hat{\range}_2^{(L)}]$ \\
\hline
$v_1 \geq v_2 \geq \tau$ & $0$ \\
$v_1 \geq \tau \geq v_2$ & $-4v_1v_2\tau(2v_1-v_2)\ln\frac{\tau}{v_2}$\\
&$+4v_1 v_2(\tau)^2+\frac{(\tau)^4}{3}+\frac{8v_2^3\tau}{3}$\\
 & $-6v_1 v_2^2\tau-4v_1^2 v_2^2$\\
& $-v_2^4+4 v_1v_2^3+4v_1^2 (\tau)^2 -2v_1(\tau)^3 $\\
$v_2 \leq v_1 \leq \tau$ & $\frac{2(\tau)^2}{3v_1}\bigg(4v_2^3+5v_1^3-9v_1 v_2^2\bigg)-(v_1-v_2)^4$\\
& $-4(\tau)^2 (2v_1-v_2)v_2\ln(\frac{v_1}{v_2})$
\end{tabular}
}
\caption{$\hat{\range}_2^{(L)}$ and its variance for independent samples.\label{indL2L:table}}
\end{table}






\section{Shared-seed sampling} \label{exsharedests:sec}
We derive estimators for $\range_p(\vecv)$($p>0$), where
$\vecv=(v_1,\ldots,v_r)$ for $r \geq 2$.
The sampling uses the same random seed $u$ for all entries.
The outcome $S(u,\vecv)$ is determined by the data $\vecv$ and
a {\em scalar} seed value $u\in (0,1]$, drawn uniformly at random:
Entry $i$ is included in $S$
if and only if $v_i \geq \tau_i u$.

We apply
our work on estimators for monotone sampling \cite{sorder:PODC2014} to 
derive two unbiased nonnegative admissible  estimators: The
\L\ estimator $\hat{\range}_p^{(L)}$ and the \U\ estimator $\hat{\range}_p^{(U)}$.
We present closed form expressions of estimators and variances when
$\mbox{\boldmath{$\tau$}}$ has all
entries equal (to the scalar $\tau$).
 The derivations easily extend to non-uniform
 $\boldsymbol{\tau}$. 

The set of data
vectors consistent with outcome $S(u,\vecv)$ is
$$S^*=\{\vecz | \forall i\in [r], i\in S\implies  z_i=v_i 
 \, , \, i\not\in S \implies z_i< \tau_i u   \}\ .$$
Observe that the sets $S^*(u,\vecz)$ are the
same for all consistent
data vectors  $\vecz\in
S^*(u,\vecv)$.
Fixing the data $\vecv$, 
the set $S^*(u,\vecv)$ is
non-decreasing with $u$, which means that the information on
the data that we can glean from the outcome can only increase when $u$
decreases.  This makes the sampling scheme {\em monotone}  in the
randomization, which allows us to apply the estimator derivations in 
\cite{sorder:PODC2014}.


\smallskip
\noindent
{\bf The lower bound function.}
The derivations use 
the {\em lower bound function} $\underline{\range}_p$, which
maps an outcome $S$ to the infimum of $\range_p$ values on
vectors that are consistent with the outcome:
$$\underline{\range}_p(S)= \inf_{\vecv\in S^*} \range_p(\vecv)
\ .$$ 
For $\range$, the lower bound is
the difference between a lower bound on the maximum entry and
an upper bound on the minimum entry.
\begin{eqnarray*}
\underline{\range}(S) & = &\max_{i\in S} v_i-\min\{\min_{i\in S} v_i ,
\min_{i\not\in S} \tau_i u\}\ .
\end{eqnarray*}
The lower bound on $\range_p$ is the $p$th power of the respective bound
on $\range$, that is,
$\underline{\range}_p(S)=\underline{\range}(S)^p$.  For $S(u,\vecv)$, we use
the notation $\underline{\range}_p(S(u,\vecv))\equiv
\underline{\range}_p(u,\vecv)$.
 For all-entries-equal $\mbox{\boldmath{$\tau$}}$:\\
\smallskip
\noindent
 \begin{tabular}{l|l|l}
\hline
condition & $|S|$ & $\underline{\range}(S)$ \\ 
\hline
$u  > \frac{\max(\vecv)}{\tau}$ & $0$ & $0$ \\ 
$\frac{\max(\vecv)}{\tau} \geq u \geq \frac{\min(\vecv)}{\tau}$ & $1\ldots r-1$ & $\max(\vecv)-u\tau$ \\ 
$u< \frac{\min(\vecv)}{\tau}$ & $r$ & $\range(\vecv)$ 
\end{tabular}

\ignore{
\begin{figure*}[htbp]
\center
\begin{tabular}{ccc}
\ifpdf
\includegraphics[width=0.3\textwidth]{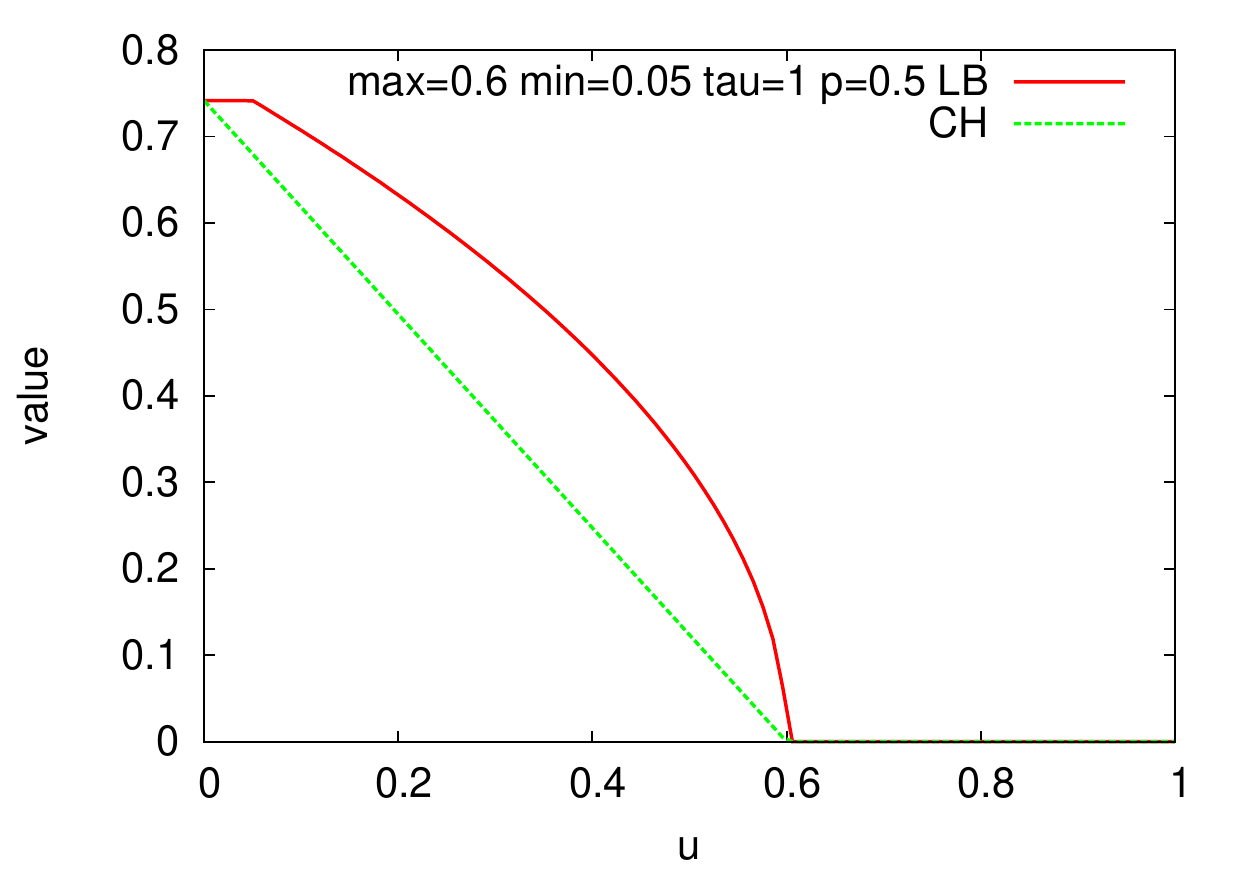} &
\includegraphics[width=0.3\textwidth]{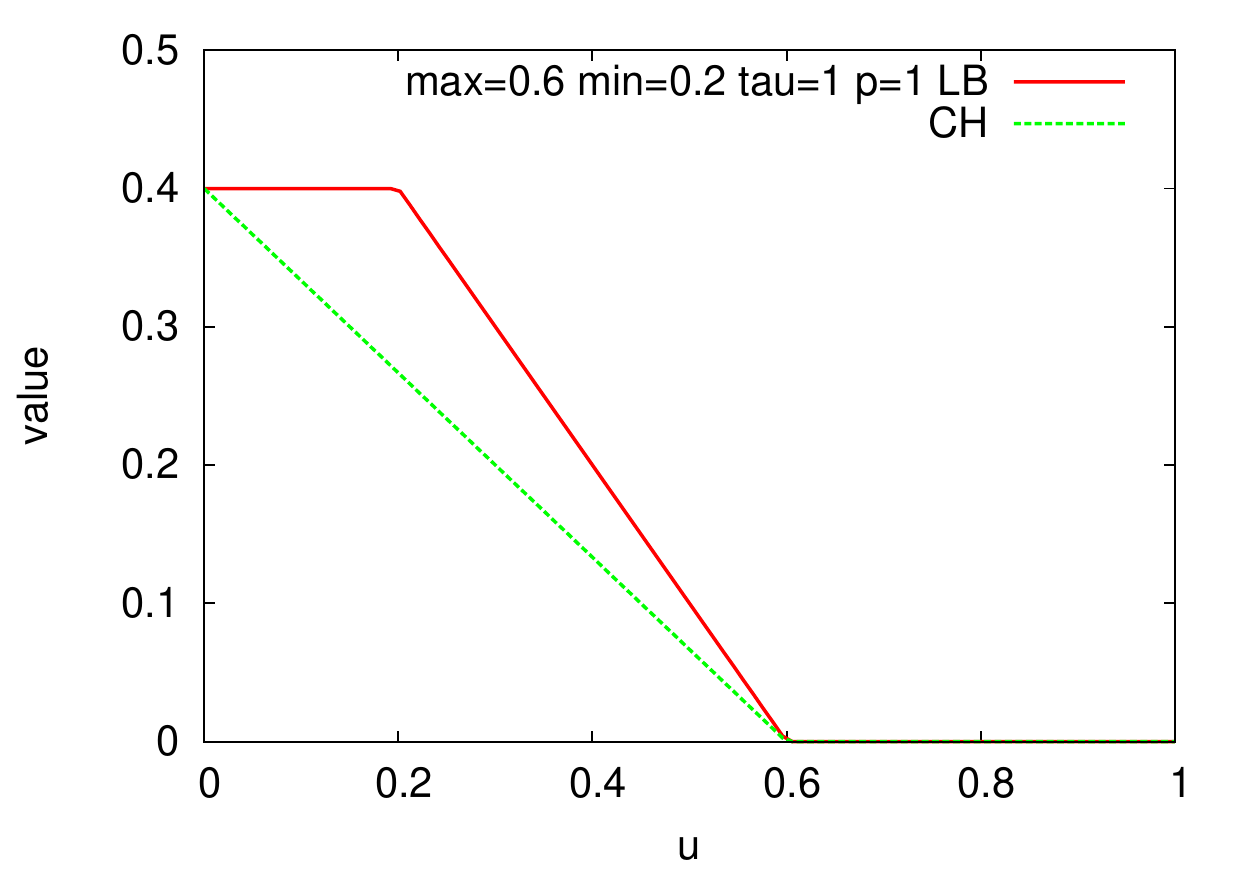} &
\includegraphics[width=0.3\textwidth]{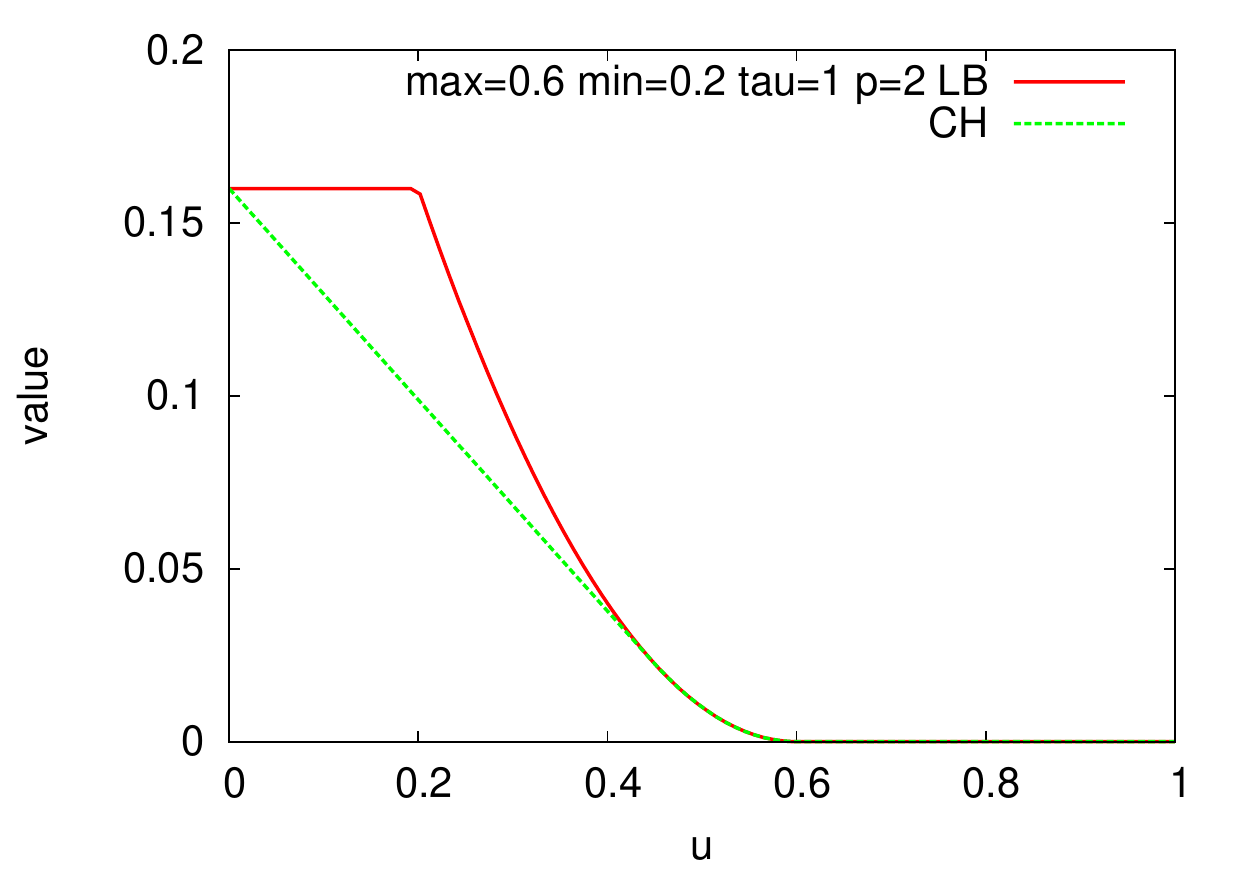}\\
  \includegraphics[width=0.3\textwidth]{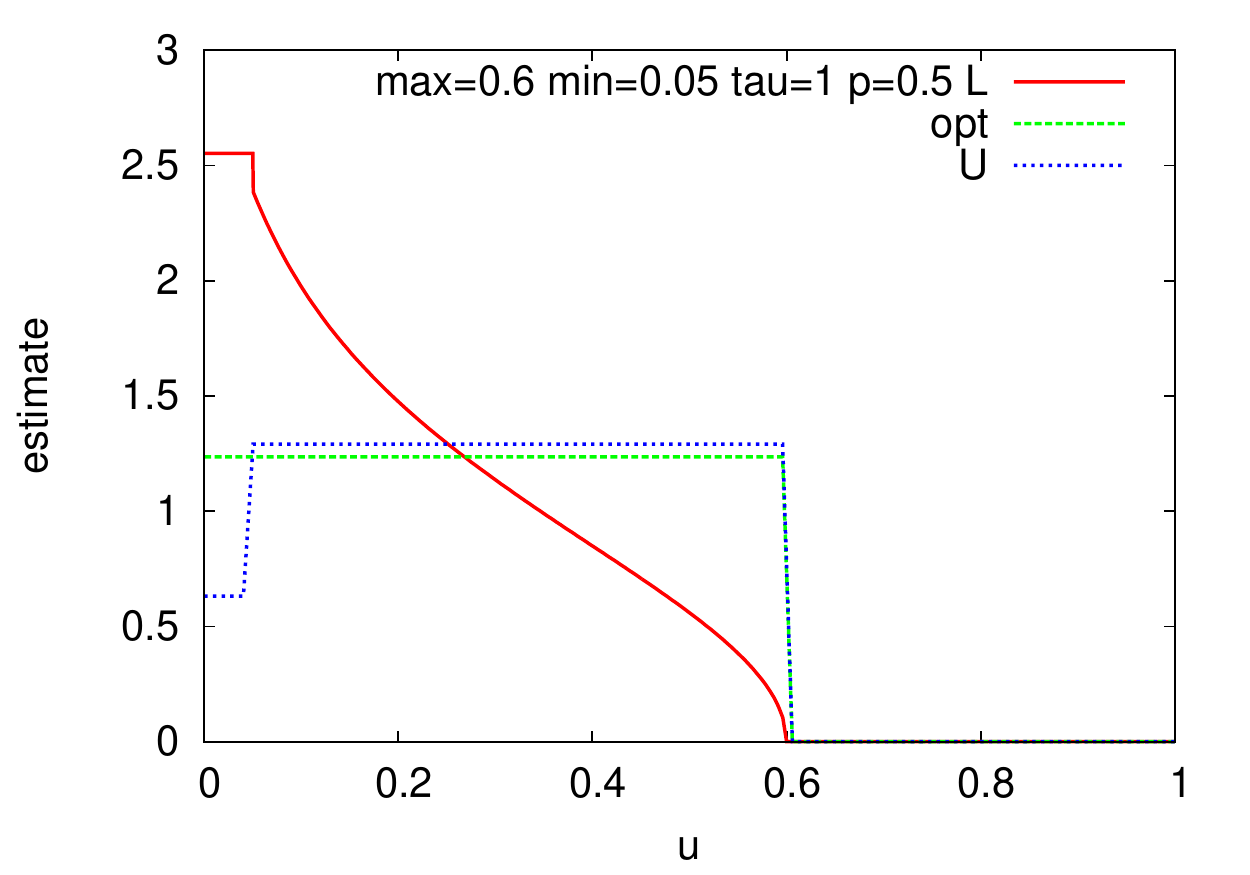} &
 \includegraphics[width=0.3\textwidth]{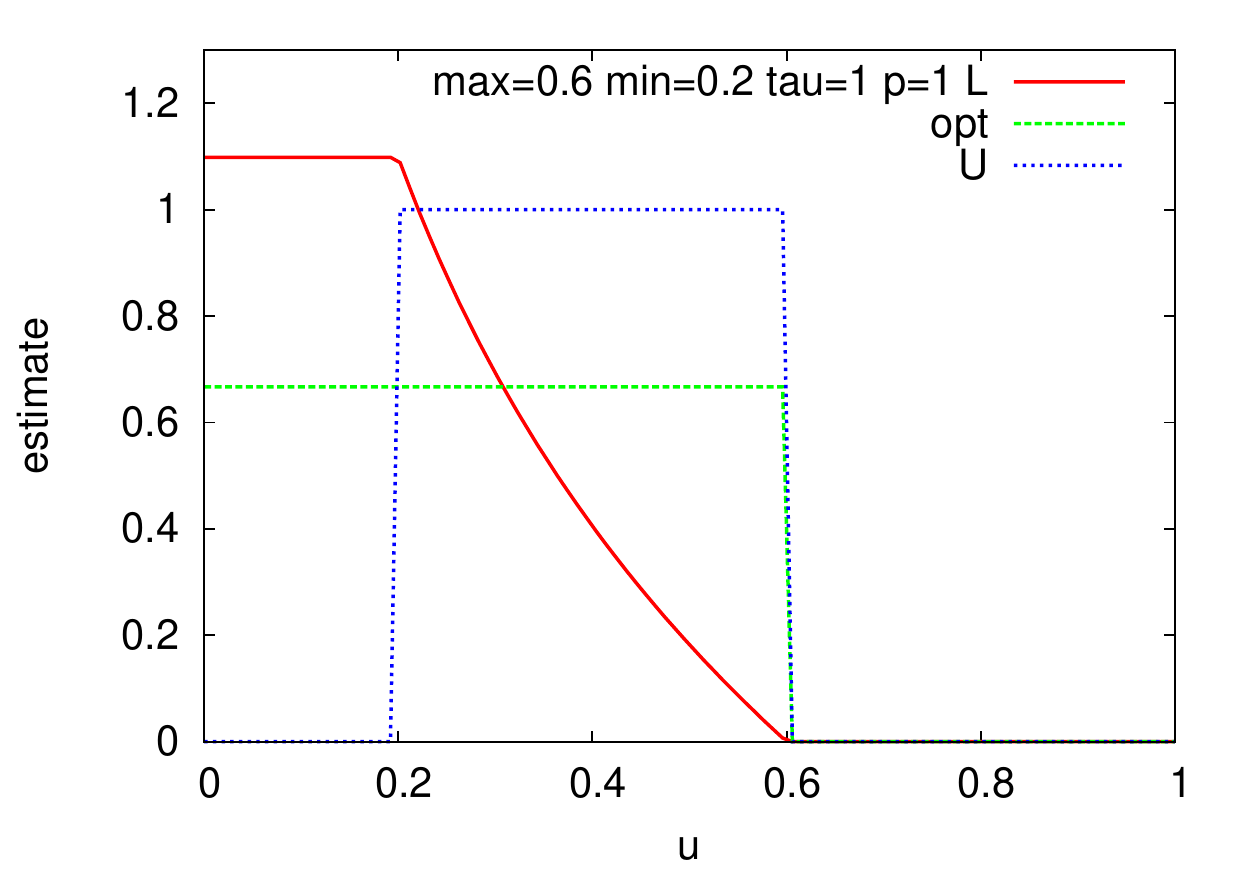} &
  \includegraphics[width=0.3\textwidth]{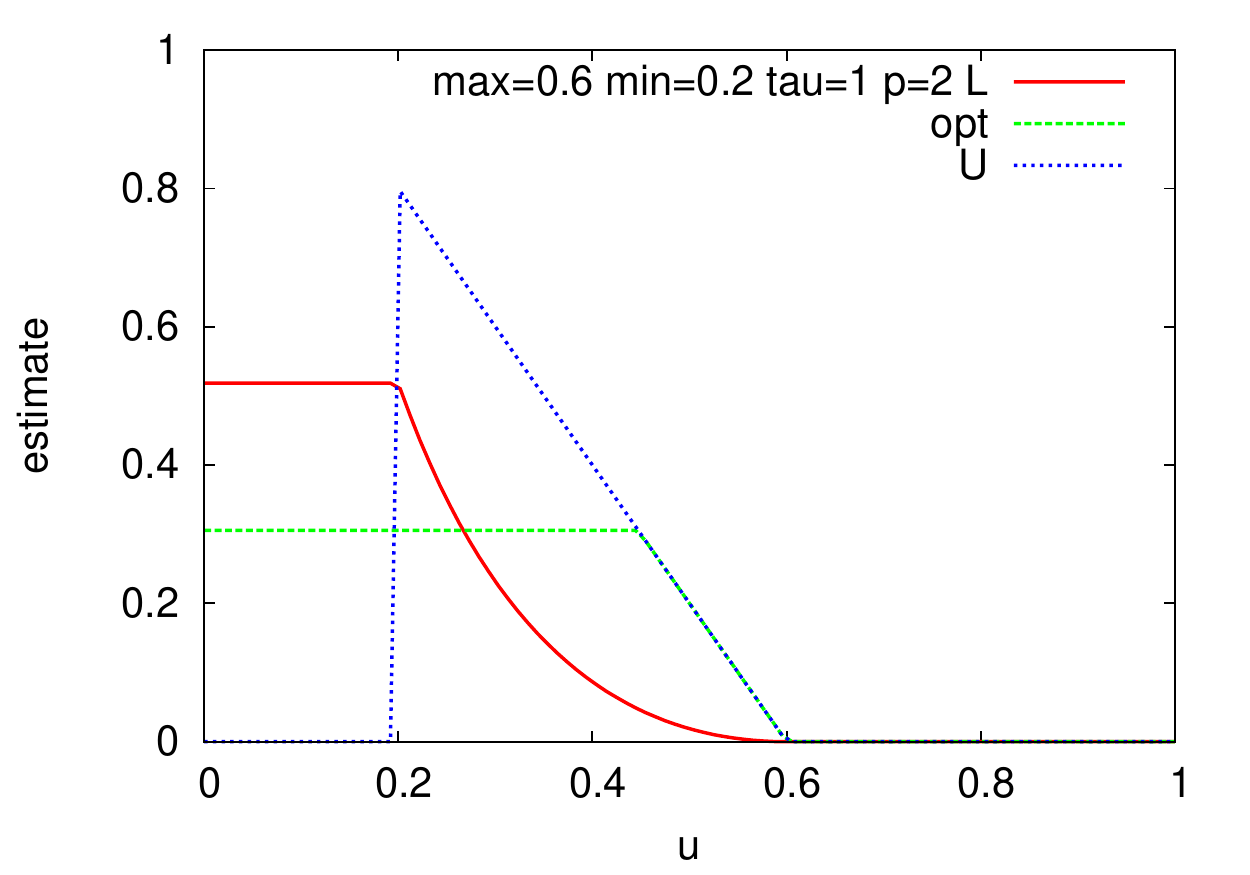}\\
\else
\epsfig{figure=phalf_LB_CH.eps,width=0.3\textwidth} &
\epsfig{figure=pone_LB_CH.eps,width=0.3\textwidth} &
\epsfig{figure=ptwo_LB_CH.eps,width=0.3\textwidth} \\
 \epsfig{figure=phalf_OLU_ests.eps,width=0.3\textwidth} &
 \epsfig{figure=pone_OLU_ests.eps,width=0.3\textwidth} &
   \epsfig{figure=ptwo_OLU_ests.eps,width=0.3\textwidth} \\
\fi
\end{tabular}
\caption{Top: The lower bound function and corresponding lower hull for example vectors and $p\in \{0.5,1,2\}$. Bottom: 
  the corresponding optimal, \L, and \U\ estimates on outcomes consistent
  with the vector. \label{optlu:fig}}
\end{figure*}
}
\begin{figure}[htbp]
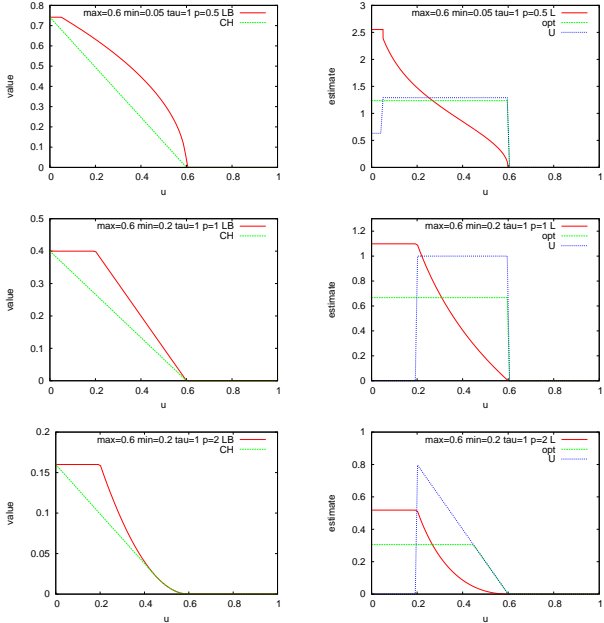

\center
\begin{tabular}{cc}
\ifpdf
\includegraphics[width=0.22\textwidth]{phalf_LB_CH} &   \includegraphics[width=0.22\textwidth]{phalf_OLU_ests} \\
\includegraphics[width=0.22\textwidth]{pone_LB_CH} &  \includegraphics[width=0.22\textwidth]{pone_OLU_ests} \\
\includegraphics[width=0.22\textwidth]{ptwo_LB_CH} &
  \includegraphics[width=0.22\textwidth]{ptwo_OLU_ests}\\
\else
\epsfig{figure=phalf_LB_CH.eps,width=0.22\textwidth} &  \epsfig{figure=phalf_OLU_ests.eps,width=0.22\textwidth} \\
\epsfig{figure=pone_LB_CH.eps,width=0.22\textwidth} &  \epsfig{figure=pone_OLU_ests.eps,width=0.22\textwidth} \\
\epsfig{figure=ptwo_LB_CH.eps,width=0.22\textwidth} &
   \epsfig{figure=ptwo_OLU_ests.eps,width=0.22\textwidth} \\
\fi
\end{tabular}
\caption{Left: The lower bound function and corresponding lower hull for example vectors and $p\in \{0.5,1,2\}$. Right: 
  the corresponding optimal, \L, and \U\ estimates on outcomes consistent
  with the vector. \label{optlu:fig}}
\end{figure}

\smallskip
\noindent
{\bf The \L\ estimator.}
The \L\ estimator \cite{sorder:PODC2014}  is defined for any monotone estimation problem for
which an unbiased and nonnegative estimator exists.  This estimator
is specified as a function of the corresponding lower bound
function. For $\range_p$ we have
\begin{align} 
\forall S(\zeta,\vecv),\,
\hat{\range}_p^{(L)}(S) &= \frac{\underline{\range}_p(\zeta,\vecv)}{\zeta}
 -\int_\zeta^1 \frac{\underline{\range}_p(u,\vecv)}{u^2}du \ .\label{shortLB}
\end{align}

 From \cite{sorder:PODC2014}, we know that $\hat{\range}_p^{(L)}$ has
 the following properties:
\begin{trivlist}
\item
$\bullet$ It is nonnegative and unbiased.
\item $\bullet$
It is the unique (up to equivalence) admissible unbiased nonnegative
monotone estimator,
 meaning that  the estimate is non-decreasing with $u$.
\item $\bullet$
It is  $\prec^+$-optimal with respect to the partial order $\prec$
$$\vecv \prec \vecz \iff \range_p(\vecv) < \range_p(\vecz)\
.$$
$\prec^+$-optimality with respect to this particular order means that any estimator with a strictly lower variance
for a data vector must have strictly higher variance on some vector
with a smaller range  -- this means
that the \L\ estimator ``prioritizes''
data where the range (or difference when aggregated) is
small.  
\end{trivlist}
The \L\ estimator has finite variances when the monotone estimation
problem admits a nonnegative estimator with finite variances.
It is also 4-{\em competitive} in terms of variance \cite{CKsharedseed:2012}, meaning that for any data vector, the
ratio of the expectation of the square to the minimum one possible for
the data via an unbiased nonnegative estimator is at most $4$
$$\forall \vecv, \E_{S\sim \mathcal{S}_{\vecv}}
[\hat{\range}_p^{(L)}(S)^2]  \leq   4 \E_{S\sim \mathcal{S}_{\vecv}}
[\hat{\range}_p^{(\vecv)}(S)^2] \ ,$$
where $\hat{\range}_p^{(\vecv)}(S)$ is a nonnegative unbiased  estimator which minimizes
the variance for $\vecv$ (We will present a construction of this estimator).
Competitiveness,  is a strong property that means that
for {\em all} data vectors,  the variance under the \L\ estimator
 is not too far off 
the minimum possible variance for that vector by a nonnegative
unbiased estimator.

 For our sampling scheme with all entries equal $\tau$, we define
$\max(\vecv) \equiv \max_i v_i$  (which is available from $S$ whenever
$|S|>0$) and 
$v_{\min}=\min(\vecv)$ if $|S|=r$ and $v_{\min}=u \tau$ otherwise,
which is also always available from $S$, the estimator is
\begin{equation}\label{rangedest:eq}
\hat{\range}_p^{(L)}(S)=
\left\{\begin{array}{ll}
|S|=0 \mbox{:}  \, & 0 \\
|S|\geq 1 \mbox{:}  \, & (\max(\vecv)-v_{\min})^p\max\{1,\frac{\tau}{v_{\min}}\} -  \\ & 
\int_{\min\{1,\frac{v_{\min}}{\tau}\}}^{\min\{1,\frac{\max(\vecv)}{\tau}\}} \frac{(\max(\vecv)-x\tau)^p}{x^2} dx
\end{array}\right.
\end{equation}
Estimators and variance for $\range$ and $\range_2$ 
are provided in Tables~\ref{rangeest:tab} and~\ref{range2est:tab}.
We also compute a tight ratio on variance competitiveness for $p=1,2$:
\begin{lemma} \label{varratio:lemma}
\begin{eqnarray}
\forall \vecv,\  \E_{S\sim
  \mathcal{S}_{\vecv}}[\hat{\range}^{(L)}(S)^2] &\leq& 2 \E_{S\sim
  \mathcal{S}_{\vecv}}[\hat{\range}^{(\vecv)}(S)^2]  \label{comp1}\\
\forall \vecv,\  \E_{S\sim
  \mathcal{S}_{\vecv}}[\hat{\range}_2^{(L)}(S)^2] &\leq& 2.5 \E_{S\sim
  \mathcal{S}_{\vecv}}[\hat{\range}_2^{(\vecv)}(S)^2]   \label{comp2}
\end{eqnarray}
\end{lemma}

\smallskip
\noindent
{\bf The \U\ estimator.}
The \U\ estimator \cite{sorder:PODC2014} is the solution of the
integral equation
\begin{align} \label{Udefeq:eq}
\forall S(\zeta,\vecv),\ \hat{\range_p}(\zeta,\vecv) = 
&\sup_{\vecz\in S^*}\inf_{0\leq \eta < \zeta}
\frac{\underline{\range}_p(\eta,\vecz)-\int_{\zeta}^1
  \hat{\range_p}(u,\vecv) du}{\zeta-\eta} \nonumber
 \end{align}
From \cite{sorder:PODC2014}, we know that $\hat{\range}_p^{(U)}$
 has the following properties:
\begin{trivlist}
\item $\bullet$
It is nonnegative and unbiased.
\item $\bullet$
It is  $\prec^+$-optimal with respect to the partial order $\prec$
$$\vecv \prec \vecz \iff \range_p(\vecv) > \range_p(\vecz)\
.$$ 
This means that the \U\ estimator ``prioritizes''
data where the range (or difference when aggregated) is
large.
In particular, it is the nonnegative unbiased estimator with minimum
variance on data with $\min(\vecv)=0$.
\end{trivlist}

The solution 
for all-entries equal $\tau$ is provided as
Algorithm~\ref{rangedUshared:alg} \notinproc{(see
 Appendix~\ref{deriveRGU} for calculation).}\onlyinproc{ (details are
 omitted due to space limitations).}
The estimator is admissible and has finite variances for all data vectors.



\onlyinproc{Expressions for the variance of $\hat{\range}_p^{(U)}$ for
  $p=1,2$ are omitted due to space limitations.}
\notinproc{The estimator $\hat{\range}_p^{(U)}$ and its variance for $p=1,2$
are provided in Tables~\ref{rg1Uest:tab} and~\ref{rg2Uest:tab}
(See Appendix~\ref{range12Ucalc:sec} for details).}

\smallskip
\noindent
{\bf $\vecv$-optimality.}
We say that an estimator is {\em $\vecv$-optimal} (for data $\vecv$),  if amongst all 
estimators that are 
nonnegative and unbiased for all data, it has 
the minimum possible variance when the data is $\vecv$. 
In order to measure the competitiveness of our estimators, as in Lemma \ref{varratio:lemma},
we derive an expression for  the {\em $\vecv$-optimal 
  estimate values} $\hat{\range_p}^{(\vecv)}$.
It turns out that the values assumed by a $\vecv$-optimal estimator on 
outcomes consistent with $\vecv$ are unique (almost everywhere on
$\mathcal{S}_{\vecv}$ \cite{CKsharedseed:2012}. 
Note that there is no single estimator that is $\vecv$-optimal for all 
$\vecv$, that is, there is no uniform minimum variance unbiased
nonnegative estimator for $\range_p$.
Therefore, the $\vecv$-optimal estimates for all possible values of $\vecv$ can not be combined into a 
single estimator.

  We now obtain an explicit representation of $\hat{\range}_p^{(\vecv)}$.
We  use the notation 
 $H_{\range_p}^{(\vecv)}(u)$ 
for the
  lower boundary of the convex hull (lower hull) of
  $\underline{\range}_p(u,\vecv)$
and the point $(1,0)$.
 This function is monotone 
non-increasing in $u$ and therefore differentiable almost everywhere.
We apply the following
 \begin{theorem} \label{uniquevopt} \cite{CKsharedseed:2012}
A nonnegative unbiased estimator $\hat{\range}_p$  minimizes
$\var_{S\sim \mathcal{S}_{\vecv}}[\hat{\range}_p]$  
$\iff$ {\em almost everywhere} on $\mathcal{S}_{\vecv}$
\begin{equation}  \label{vopt}
\hat{\range}_p^{(\vecv)}(u)= -\frac{d  H_{\range_p}^{(\vecv)}(u)}{du}\ .
\end{equation}
 \end{theorem}
The estimates \eqref{vopt} are monotone 
non-increasing in $u$.  

We can now specify  $\hat{\range}_p^{(\vecv)}$  for PPS sampling
with all-entries-equal $\tau$.
The function $\underline{\range}_p(u,\vecv)$ is
$\max\{0,\max(\vecv)-\tau\}$ for $u\geq
\frac{\max(\vecv)}{\tau}$ and equal to $\range_p(\vecv)$ for
$u\leq
\frac{\min(\vecv)}{\tau}$.  Therefore for $u\geq
\frac{\max(\vecv)}{\tau}$, 
the lower hull is 
$H_{\range_p}^{(\vecv)}(u) =0$ and $\hat{\range}_p^{(\vecv)}(u)=0$.

For $p\leq 1$, the function is concave for
$u\in [\frac{\min(\vecv)}{\tau},\frac{\max(\vecv)}{\tau}]$.
The lower hull is therefore
a linear function for $u\leq \frac{\max(\vecv)}{\tau}$: when
$\max(\vecv)\leq \tau$,  $H_{\range_p}^{(\vecv)}(u) =\range_p(\vecv)(1-u\frac{\tau}{\max(\vecv)})$
and when $\max(\vecv)\geq \tau$,  $H_{\range_p}^{(\vecv)}(u) =\range_p(\vecv)-u(\range_p(\vecv)-(\max(\vecv)-\tau)^p)$.
The $\vecv$-optimal estimates are therefore constant for $u\leq \min\{1,\frac{\max(\vecv)}{\tau}\}$:
$\hat{\range}_p^{(\vecv)}(u)=\range(\vecv)\frac{\tau}{\max(\vecv)} $ when  $\max(\vecv)\leq \tau$, and 
$\hat{\range}_p^{(\vecv)}(u)=\range_p(\vecv)-(\max(\vecv)-\tau)^p$ when $\max(\vecv)\geq \tau$.

For $p>1$, $\underline{\range}_p(u,\vecv)$ is convex for
$u\in [\frac{\min(\vecv)}{\tau},\frac{\max(\vecv)}{\tau}]$.
Geometrically, the
lower hull follows the lower bound function for $u>\alpha$, where
$\alpha$ is the point where the slope of the lower bound function is
equal to the slope of a line segment connecting the current point to
the point $(0,\range_p(\vecv))$.
For $u\leq \alpha$, the lower hull follows this line segment and is
linear.
  Formally, the point $\alpha$ is the solution
of
$$ \range_p(\vecv)= (\max(\vecv)-x\tau)^{p-1} (p\tau+\max(\vecv)-x\tau)\ .$$
If there is no solution $\alpha\in [\frac{\min(\vecv)}{\tau},
\min\{1,\frac{\max(\vecv)}{\tau}\}]$, 
we use $\alpha=\min\{1,\frac{\max(\vecv)}{\tau}\}$.
The  estimates  for $u\in [\alpha,\min\{1,\frac{\max(\vecv)}{\tau}\}]$ are
$\hat{\range}_p(u,\vecv)=-\frac{d\underline{\range}_p(u,\vecv)}{du}=p\tau(\max(\vecv)-u\tau)^{p-1}$ and for
$u\leq \alpha$, $\hat{\range}_p^{(\vecv)}(u)=\frac{\range_p(\vecv)-(\max(\vecv)-\alpha\tau)^p}{\alpha}$.

 Figure~\ref{optlu:fig} (top) illustrates
 $\underline{\range}_p(u,\vecv)$  and
the corresponding lower hull
$H_{\range_p}^{(\vecv)}$  as a function of $u$ for example vectors with
 $p\in\{0.5,1,2\}$.

Now that we expressed $\hat{\range}_p^{(\vecv)}(S)$ on all outcomes
consistent with $\vecv$,  we can
 compute (for any vector $\vecv$),
  the minimum possible variance attainable for it by an unbiased
  nonnegative estimator:
\begin{equation}  \label{voptest}
\var_{\mathcal{S}_{\vecv}}[\hat{\range}_p^{(\vecv)}]=\int_0^1 \hat{\range}_p^{(\vecv)}(u) ^2 du - \range_p(\vecv)^2\ .
\end{equation}
We use the expectation of the square 
to measure the 
``variance competitiveness'' of estimators (Since the second summand
in \eqref{voptest} is the same for all estimators).  The ratio for a
particular $\vecv$ is the expectation of the square for $\vecv$
divided by the optimal one which is
$\E_{S\sim
  \mathcal{S}_{\vecv}}[\hat{\range}_p^{(\vecv)}(S)^2]=\int_0^1
\hat{\range}_p^{(\vecv)}(u) ^2 du$.
The competitive ratio is the maximum ratio over all $\vecv$.

\ignore{  
\begin{figure*}[htbp]
\centerline{
\begin{tabular}{ccc}
\ifpdf
\includegraphics[width=0.3\textwidth]{sharedS_L1_HT_LB_var_0_25} &
\includegraphics[width=0.3\textwidth]{sharedS_L1_HT_LB_var_0_01} &
\includegraphics[width=0.3\textwidth]{sharedS_L1_LB_r_dHT_var}\\
\else
\epsfig{figure=sharedS_L1_HT_LB_var_0.25.eps,width=0.3\textwidth} &
\epsfig{figure=sharedS_L1_HT_LB_var_0.01.eps,width=0.3\textwidth} &
\epsfig{figure=sharedS_L1_LB_r_dHT_var.eps,width=0.3\textwidth} \\
\fi
(A) & (B) & (C)
\end{tabular}}
\caption{Figures (A) and (B) show the variance of $\hat{\range}^{(U)}$ and $\hat{\range}^{(L)}$, scaled by $(\tau)^2$ to be independent of $\tau$, illustrating the dependence of the variance on the ratio $\min(\vecv)/\max(\vecv) \in[0,1]$ for $\max(\vecv)/\tau \in \{0.01,0.25\}$.  Figure (C) illustrates the dependence of the ratio $\var[\hat{\range}^{(L)}]/\var[\hat{\range}^{(U)}]$ on $\min(\vecv)/\max(\vecv)$ for different values of $\max(\vecv)/\tau$. \label{dHT_LB_poisson_sharedS:fig}}
\end{figure*}
}

\notinproc{
\begin{figure*}[htbp]
\center
\begin{tabular}{ccc}
\ifpdf
\includegraphics[width=0.3\textwidth]{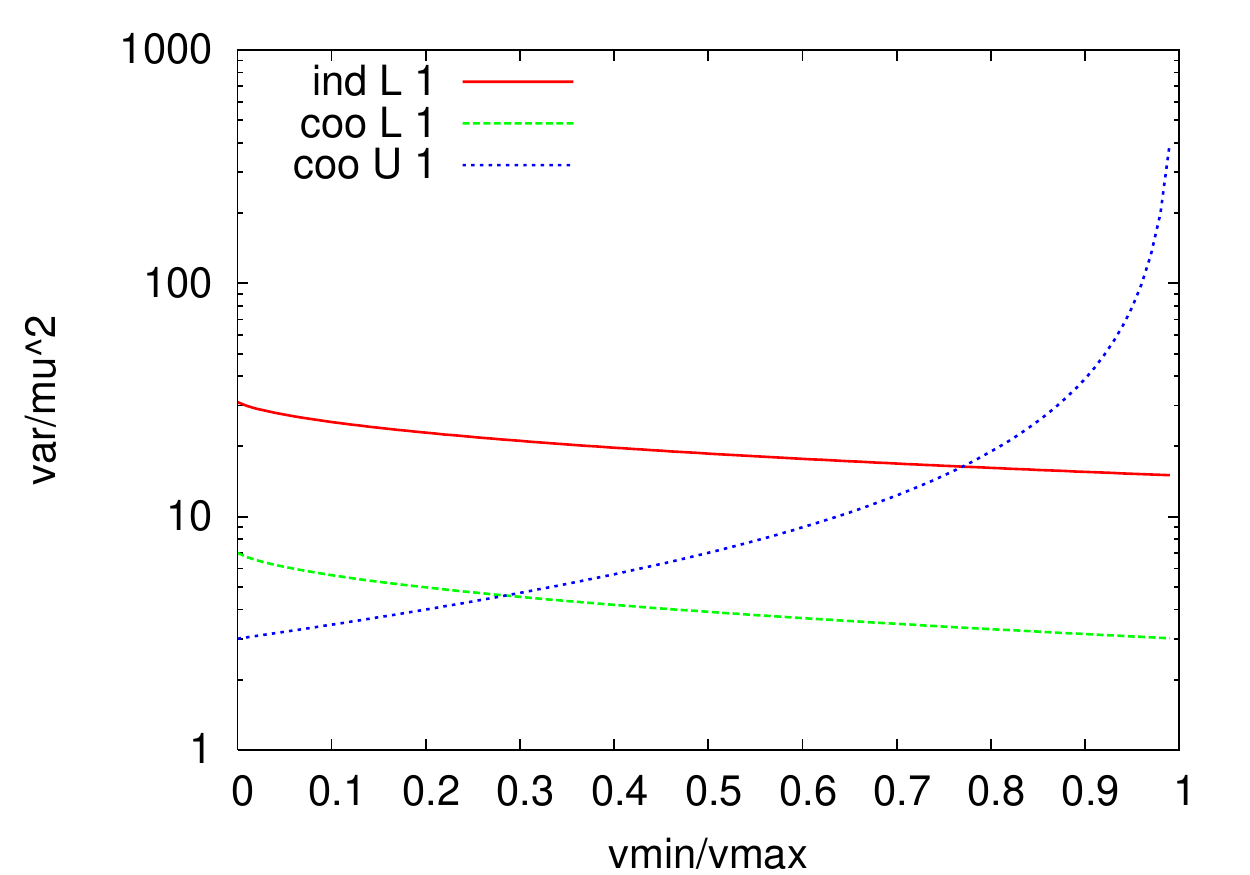} &
\includegraphics[width=0.3\textwidth]{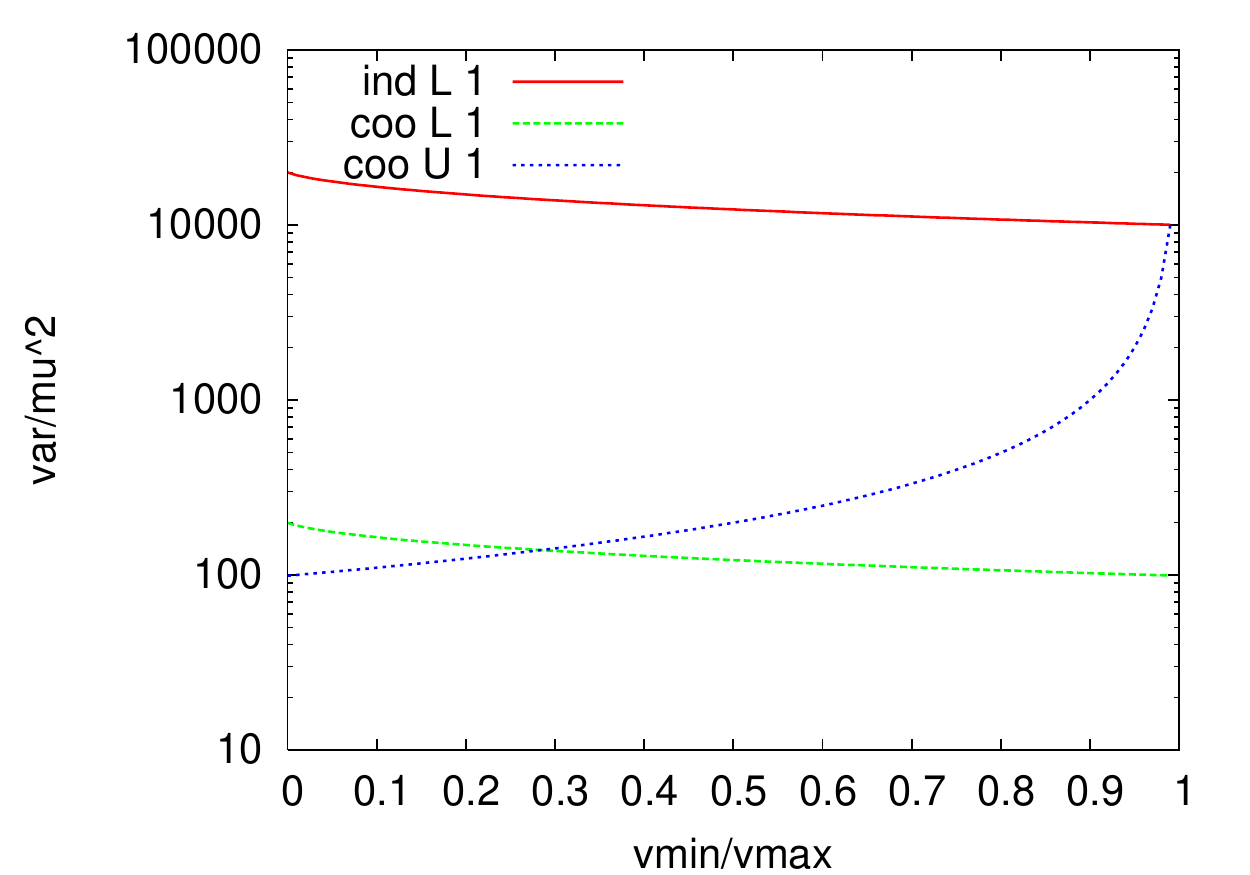} &
\includegraphics[width=0.3\textwidth]{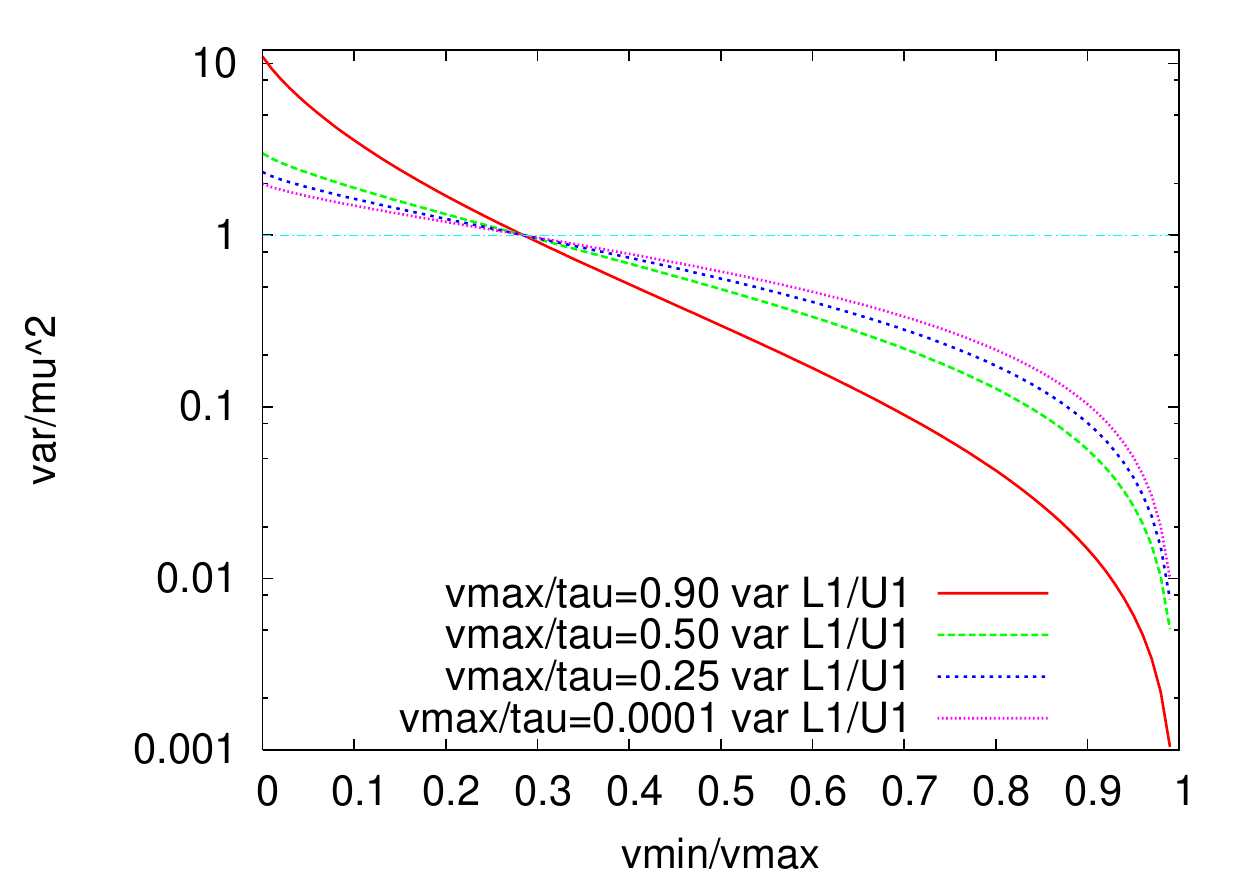}\\
\includegraphics[width=0.3\textwidth]{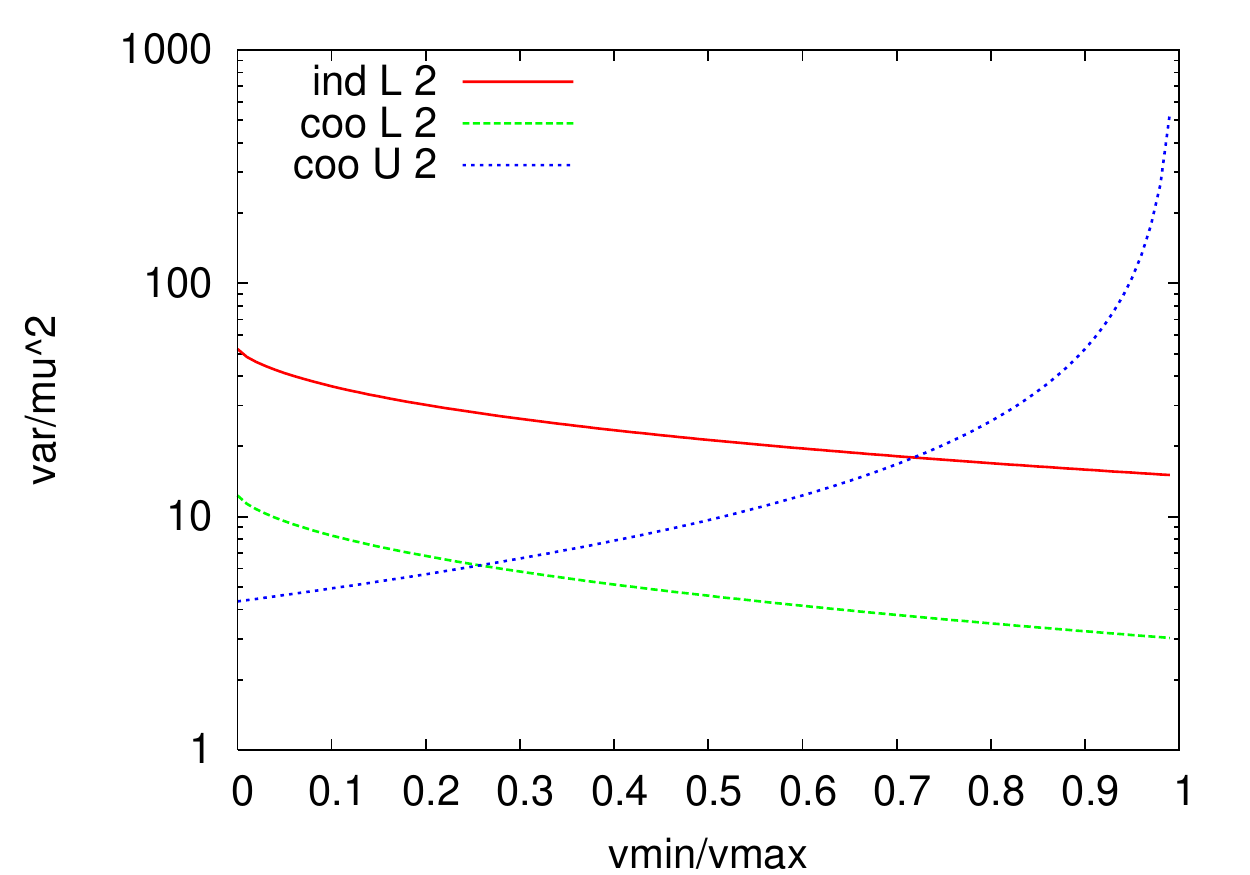} &
\includegraphics[width=0.3\textwidth]{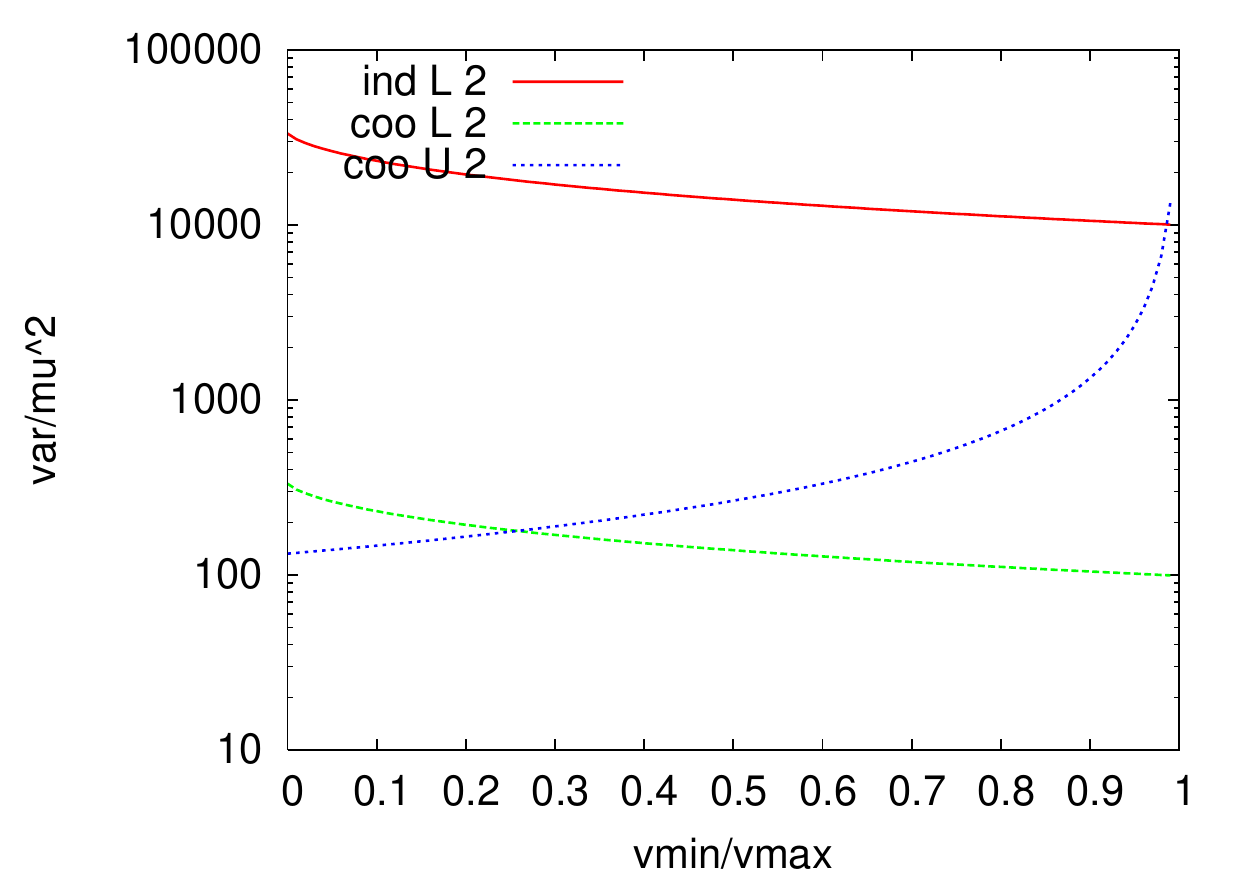} &
\includegraphics[width=0.3\textwidth]{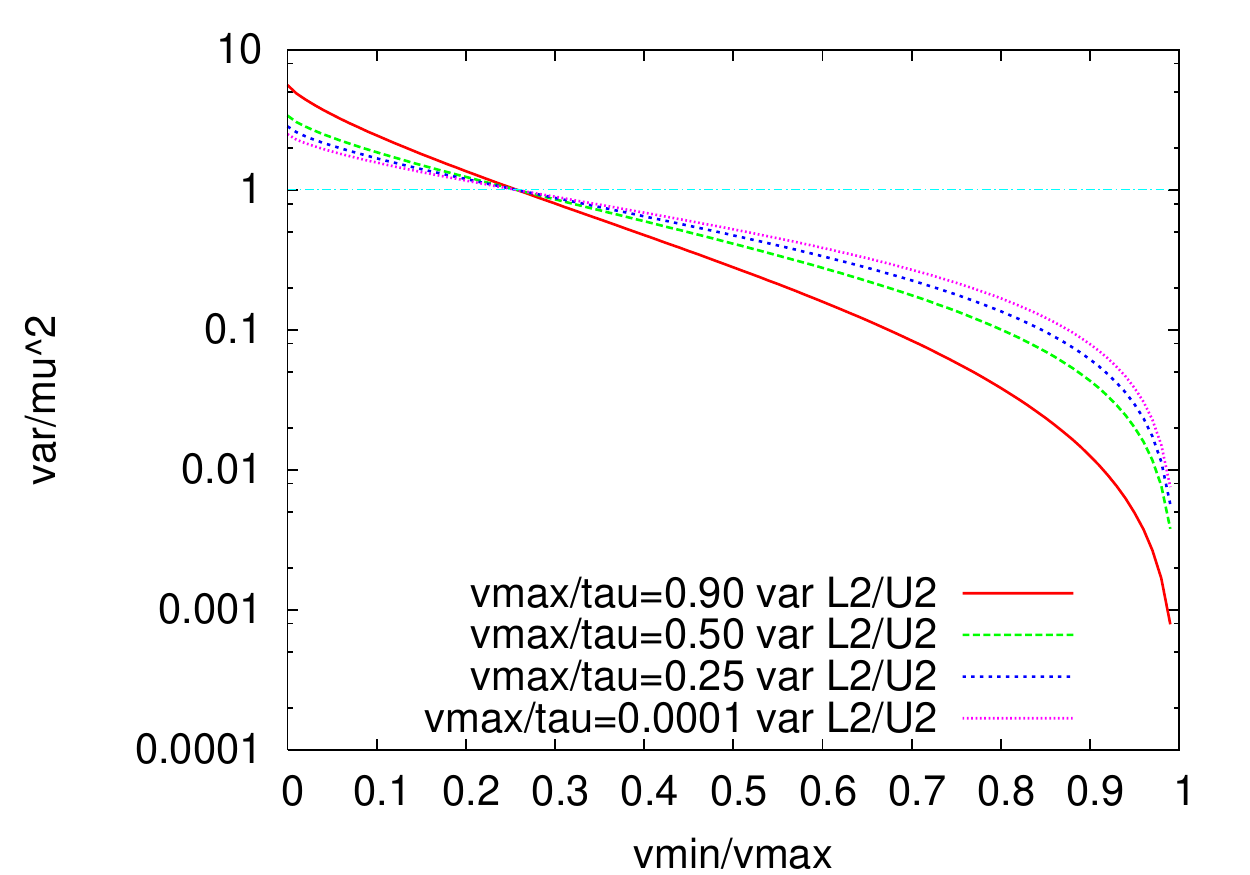}\\
\else
\epsfig{figure=T100_vmax25_ic_var1.eps,width=0.3\textwidth} &
\epsfig{figure=T100_vmax1_ic_var1.eps,width=0.3\textwidth} &
\epsfig{figure=var_ratio_cUL1.eps,width=0.3\textwidth} \\
\epsfig{figure=T100_vmax25_ic_var2.eps,width=0.3\textwidth} &
\epsfig{figure=T100_vmax1_ic_var2.eps,width=0.3\textwidth} &
\epsfig{figure=var_ratio_cUL2.eps,width=0.3\textwidth} \\
\fi
(A) & (B) & (C)
\end{tabular}
\caption{Variance (normalized by square of expectation) of $\hat{\range}_p^{(L)}$ estimator over independent samples and of $\hat{\range}_p^{(L)}$ and $\hat{\range}_p^{(U)}$ over shared-seed samples. Sampling with all-entries equal $\tau$.  (A): data with $\max(\vecv)=0.25\tau$.  (B): data with $\max(\vecv)=0.01\tau$. (C):  ratio $\var[\hat{\range}_p^{(L)}]/\var[\hat{\range}_p^{(U)}]$ for shared-seed sampling, selected ratios $\max(\vecv)/\tau$.  Sweeping $\min(\vecv)$. Top shows $p=1$, bottom is $p=2$.\label{UL_range12_var:fig}}
\end{figure*}
}

\begin{figure}[htbp]
\center
\begin{tabular}{cc}
\ifpdf
\includegraphics[width=0.22\textwidth]{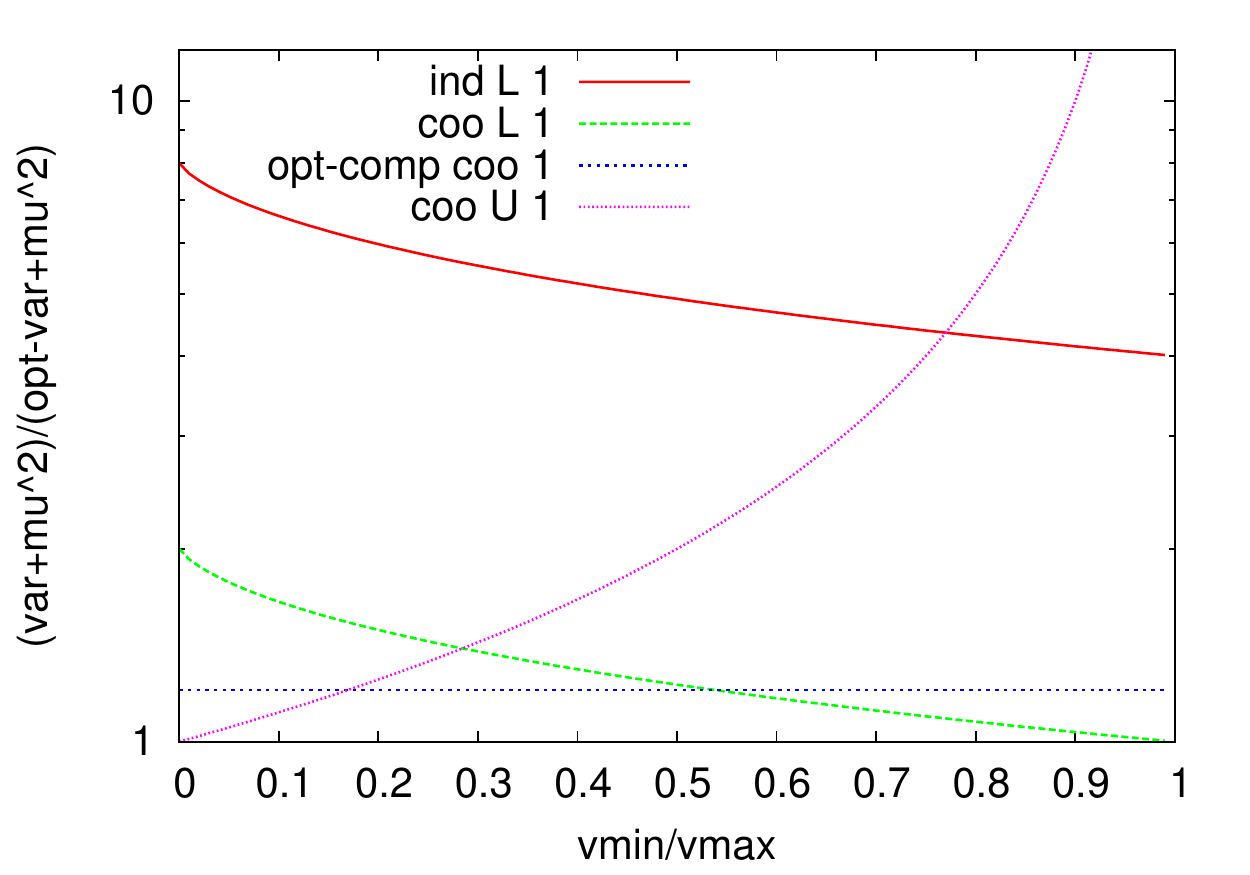} &
\includegraphics[width=0.22\textwidth]{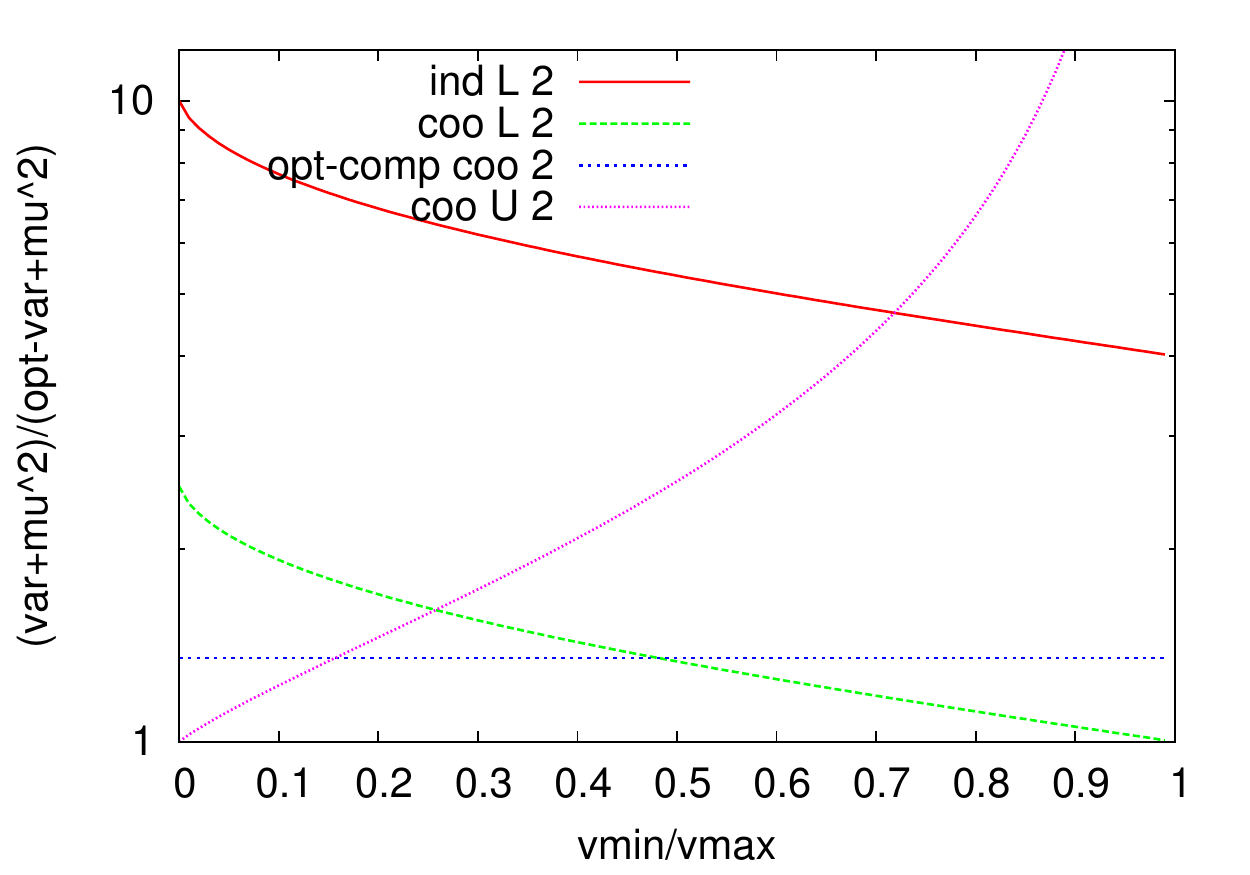} \\
\else
\epsfig{figure=T100_Esq_opt_ratio_ic_1.eps,width=0.22\textwidth} &
\epsfig{figure=T100_Esq_opt_ratio_ic_2.eps,width=0.22\textwidth} \\
\fi
$p=1$, $\frac{\max(\vecv)}{\tau}=0.25$ &  $p=2$, $\frac{\max(\vecv)}{\tau}=0.25$ 
\end{tabular}
\caption{Ratio of the expectation of the square to the minimum possible expectation of the square for the data point (over shared-seed samples), as a function of the ratio $\min(\vecv)/\max(\vecv)$.
Estimator $\hat{\range}_p^{(L)}$ over independent
samples, and $\hat{\range}_p^{(L)}$, $\hat{\range}_p^{(U)}$, and the
optimally competitive estimator 
over shared-seed samples. 
Sampling with all-entries equal $\tau$.    \label{UL_range12_Esq:fig}} 
\end{figure}

\noindent
{\bf The optimally competitive (OC) estimator.}
We used a program to compute the estimator with minimum competitive
ratio.  
The estimator was computed for an outcome that revealed $\max(\vecv)$ 
and provided an upper bound $x< \max(\vecv)$ on $\min(\vecv)$.
The domain was discretized and the estimates were computed iteratively
for decreasing $x$ so that the estimates satisfy a certain ratio $c$.
We then performed a search to find the minimum $c$ for which the computation
is successful.

\noindent
{\bf Choosing between the \L, \U, and OC estimators.}
Figure~\ref{optlu:fig} shows the $\vecv$-optimal estimates and the  \L\ and \U\
estimators for example vectors, illustrating the monotonicity of \L\ and
how the estimators relate to each other.
The estimators and their variances
 depend only on $\boldsymbol{\tau}$ and the maximum and minimum entry
 values $\max(\vecv)$ and $\min(\vecv)$.  We study the variance for
all-entries-equal $\mbox{\boldmath{$\tau$}}$
and $\max(\vecv)\leq \tau$.

Figure \ref{UL_range12_Esq:fig} shows the expectation of the square
of the \L\ estimator over independent samples and of the \L, \U, and OC estimators
over shared-seed samples.  This is as a function of the ratio of the minimum
to the maximum value in the data vector.  The expectation of the square
plotted is the ratio to the minimum possible expectation of the square 
over coordinated samples (the $\vecv$-optimum).  Recall that the $\vecv$-optimum is
not simultaneously attainable for all vectors and is used only as a
reference for variance competitiveness. 
 We can see that the \L\ estimator is nearly optimal
when the ratio is large and that the \U\ estimator is nearly optimal
when the ratio is small. The OC estimator outperforms both in the mid
range.  For the \L\ estimator, the ratio is always at most 2 (for $p=1$)
and 2.5 (for $p=2$) from the optimum whereas the \U\ estimator can
have large ratios.  The OC estimator has optimal worst-case ratios of
1.204 (for $p=1$) and 1.35 (for $p=2$), but the ratio is the same
across the range for all data vectors with $\min(\vecv) < \max(\vecv)
\leq \tau$.

We study the variance as a function of
$\frac{\min(\vecv)}{\max(\vecv)}$.
The variance is $0$ when $\range(\vecv)=0$ (ratio is $1$). Otherwise,
it is lower for
$\hat{\range}_p^{(U)}$ when the ratio is sufficiently small. The
threshold point is  $\phi_p$ which satisfies
 $$\var_{\mathcal{S}_{\vecv}}[\hat{\range}_p^{(U)}] < \var_{\mathcal{S}_{\vecv}}[\hat{\range}_p^{(L)}] \iff  \frac{\min(\vecv)}{\max(\vecv)} < \phi_p\ .$$
 For $p=1$, $\phi_1\approx 0.285$ (is the solution of the equality
$(1-x)/(2x) = \ln(1/x)$).  For $p=2$, $\phi_2\approx 0.258$.

 This suggests selecting an estimator according to expected
characteristics of the data. If typically
$\range{(\bf v)} > (1-\phi_p) \max{(\bf v)}$, we choose
$\hat{\range}_p^{(U)}$. If typically
$\range{(\bf v)} < (1-\phi_p) \max{(\bf v)}$ we choose
 $\hat{\range}_p^{(L)}$, and otherwise, we choose the OC estimator.

\notinproc{
The variance of the \L\ estimator over independent samples and of the \L\
and \U\ estimators over shared-seed samples is illustrated in
Figure~\ref{UL_range12_var:fig}.
The figure also illustrates the
relation between the variance of the shared-seed \L\ and \U\ estimators.
When $\max(\vecv)/\tau \ll 1$ (which we expect to be a prevailing
scenario), 
$\var[\hat{\range}^{(L)}]$ is nearly at most $2$ times
$\var[\hat{\range}^{(U)}]$
but as $\min(\vecv)\rightarrow \max(\vecv)$,  the ratio
$\var[\hat{\range}^{(U)}]/\var[\hat{\range}^{(L)}]$ is not bounded.
When $\max(\vecv)/\tau$ is close to $1$,  the variance of the \U\
estimator is close to $0$, and
$\var[\hat{\range}^{(L)}]/\var[\hat{\range}^{(U)}]$ is not bounded.
Interestingly, for $p=2$, the variance of the \U\ estimator is always at least
$\frac{4}{3}\range_4(\vecv)$, and thus, using \eqref{comp2},  the variance of the \L\
estimator is at most $4.4$ times the variance of the $\vecv$-optimal
(and thus of the \U\ estimator).
}

\section{Extensions}

The  {\em one-sided distance}, which isolates
the growth or decline components, is defined as
$L^p_{p+}(H)=\sum_{h\in H}
\range_{p+}(\vecv(h))$, where 
$\range_{p+}(\vecv)=\max\{0,v_1-v_2\}^p$.  
We can use
any of our $\range_p$ estimators $\hat{\range}_p(S)$ to 
estimate $\range_{p+}$ as follows:
If $S^*$ includes $\vecv$ such that $v_1 \leq v_2$ then
$\hat{\range}_{p+}(S)=0$.  Otherwise, 
$\hat{\range}_{p+}(S)= \hat{\range}_p(S)$.
We can symetrically define
$L^p_{p-}$ and $\range_{p-}$ and $\hat{\range}_{p-}$.

 Our derivations can be extended to other sampling schemes.  
One such extension is to weighted sampling without replacement (PPSWR), which is 
bottom-$k$ sampling with priorities 
$r_{ih}=  -\frac{\ln 
  (1-u_{ih})}{v_{ih}}$ \cite{Rosen1997a,Rosen1972:successive,bottomk07:ds,bottomk:VLDB2008}. 

\section{Related work}

Prior to our work, the only  
distance estimator we are aware of which obtained  good estimate with 
small fraction of data sampled is for $L_1$ over coordinated 
samples.  This estimator uses the relation $|v_1-v_2|=\max\{v_1,v_2\}-\min\{v_1,v_2\}$
to obtain an indirect estimate as the difference of two inverse 
probability estimates for the maximum and minimum
~\cite{multiw:VLDB2009}. 
Our \U\ estimator 
for $p=1$ is a strengthening of this $L_1$ estimator. 

Distance estimation over {\em unweighted} coordinated 
\cite{LiChurchHastie:NIPS2006} or independent \cite{CK:pods11}
 sampling is a much 
simpler problem. 
With unweighted sampling,
the inclusion probability of positive entries is independent of their 
weight.  
When carefully implemented, we can use 
inverse-probability estimates, which as discussed in the introduction, do not work with 
weighted sampling.   
An unweighted sample, however, is 
much less informative for its size when the data is skewed, as ``heavy 
hitters'' can be easily missed out. 
The estimators we develop here are applicable with, and take 
advantage, of weighted sampling.

Distance estimation of vectors (each instance in our terminology 
is presented as a vector of key values) 
was extensively studied using {\em linear sketches}, which are 
random linear projections, e.g.~\cite{IndykMotwani:stoc98,BGS:CIKM12,ams99,DCL:sigir08}. 
Random projections have the property that the difference vector of sketches 
is the sketch of the difference of the two vectors.  This means they 
are tailored for 
distance-like queries which aggregate over functions of the coordinate-wise 
differences. 
In particular, with linear sketches we can accurately estimate 
distances that are very small relative to the input vectors norms, whereas 
even with weighted sampling, accuracy depends 
on the relation of the distance to the vectors norms. 
A significant disadvantage of linear sketches, however, is lack of
flexibility:
Unlike samples, they  do not support domain (selection) 
queries that are specified after the summary structure is computed and
can only estimate distance between the full vectors.  Moreover, each 
sketch is tailored for a specific metric, such as $L_2^2$.  Lastly, linear sketches 
are not suitable  for estimating one-sided distances.  Moreover,
linear sketches have size which
often depends (poly) logarithmically on the number of keys.
To summarize, the two techniques, sampling and linear sketches, have 
different advantages.  Sampling provides much greater
flexibility in terms of supported queries
and often admits a smaller summary structure.

\section{Experimental Evaluation} \label{dataeval:sec}

\begin{table*}
\center
{\small
\begin{tabular}{l|rrr|rll|lll}
data & \# keys &  p1\% & p2\% & $\sum_{ih} v_{ih}$ & p1\% & p2\% & $L_1/\sum_{ih} v_{ih}$ & $L_{1+}/\sum_{ih} v_{ih}$ & $L_{1-}/\sum_{ih} v_{ih}$ \\
\hline
destIP & $3.8\times 10^4$ & 65\% & 65\% & $1.1\times 10^6$ & 49\% & 51\% & 0.36 & 0.19 & 0.18 \\
Server & 2.7$\times 10^5$ & 53\% & 56\% & $2.9\times 10^6$ & 50\% & 50\% & 0.75  & 0.38 & 0.37 \\
Surnames & 1.9$\times 10^4$ & 100\% & 100\% & $8.9\times 10^7$ & 48.6\% & 51.4\% & 0.094 & 0.0617 & 0.0327  \\
OSPD8 & 7.5$\times 10^5$ & 99\% & 99\% & 1.57$\times 10^{10}$ & 46.8\% & 53.2\% & 0.0826 & 0.0727 & 0.0099
\end{tabular}
}
\caption{Datasets with subset selections. Table shows total number of
  distinct keys $H$
  satisfying selection predicate that had a positive value in at least
  one of the two instances, corresponding percentage in each instance,
  total sum of values $\sum_{ih} v_{ih}$, and fraction (shown as
  percentage), sum  in each instance $i=1,2$: $\frac{\sum_{h} v_{ih}}{\sum_{jh} v_{jh}}$, and normalized $L_1$, $L_{1+}$ and $L_{1-}$ distances. \label{dataset:tab}}
\end{table*}

We selected several datasets that have the form of values assigned to 
a set of keys, on two instances, and natural 
selection predicates.  We consider the 
$L_1$ and $L^2_2$ distances on keys satisfying these predicates.   
Properties of the data sets with selections
are  summarized below and in Table~\ref{dataset:tab}. 

\begin{trivlist}
\item $\bullet$ 
{\bf destIP} (IP packet traces):  keys: (anonymized) IP destination addresses.  value: the number of IP flows to this destination IP. 
Instances: two consecutive time periods.  Selection: all IP 
destination addresses in a certain subnetwork.


\item $\bullet$
{\bf Server} (WWW activity logs):
Keys: (anonymized) source IP address and Web site pairs.  value: the number of HTTP 
requests issued to the Web site from this address. Instances: two 
consecutive time periods.   Selection:  A particular Web site.   


\item $\bullet$ 
{\bf Surnames and OSPD8:}
keys:  all words (terms). 
value: the number of occurrences of the term in 
English books digitized by Google and published within the time period~\cite{googlebooks:science2011}. 
Instances: the years 2007 and 2008. 
 Surnames selection:  the $18.5\times 10^3$ most common surnames in the US. 
 OSPD8 selection:  the  $7.5\times 10^4$ 8 letter words that appear in the Official 
Scrabble Players Dictionary (OSPD). 




\end{trivlist}

Each of the two 
instances were Poisson PPS~\cite{Hajekbook1981} sampled (see
Section \ref{prelim:sec})
with different sampling threshold $T$, to obtain a range of sample sizes. 
We used both coordinated (shared-seed) and independent sampling of the 
two instances.

We study the  quality of the
$L^p_p$ estimates obtained from our $\range_p$ estimators.
We estimate
$L^p_p=\sum_{h\in H} \range_p(\vecv(h))$ as the sum over selected keys $H$ of $\range_p$ estimates:
$\hat{L}^p_p = \sum_{h\in H} \hat{\range}_p(\vecv(h))$. 
We consider the estimator
$\hat{\range_p}^{(L)}$ for independent samples (Section \ref{ind:sec}) and the 
estimators $\hat{\range_p}^{(L)}$ and 
$\hat{\range_p}^{(U)}$ for coordinated samples (Section \ref{exsharedests:sec}).
To apply the estimators, we apply the selection predicate to sampled
keys to identify all the ones satisfying the predicate. The estimators
are then computed for
keys that are sampled in at least one instance (the estimate is $0$
for keys that are not sampled in any instance and do not need to be
explicitly computed).

Since all our
$\range_p$ estimators are unbiased and nonnegative, so is the
corresponding sum estimate $\hat{L}^p_p$.
 The variance is additive and is
$\sum_{h\in H}\var_{\mathcal{S}_{\vecv(h)}}[\hat{\range_p}]$.  We
measure the performance of the estimators using  
the variance normalized by the square of the expectation, which is
the squared coefficient of variation 
$\mbox{CV}^2(\hat{L}^p_p)=\frac{\sum_{h\in H}\var_{\mathcal{S}_{\vecv(h)}}[\hat{\range_p}]}{(\sum_{h\in H} \range_p(\vecv(h)))^2}$.
\notinproc{This is the same as the normalized mean squared error (MSE) for our
unbiased estimators.}

Figure~\ref{datasets_d12:fig} shows
the CV$^2$ of our $L^p_p$ estimators ($p=1,2$)
 as a function of the sampled fraction of the dataset.
We can see qualitatively, that all estimators, even over independent samples,
 are satisfactory, in
that the CV is small for a sample that is a small fraction of the full
data set.
The estimator $\hat{\range}_p^{(L)}$ over coordinated shared-seed
samples outperforms, by orders of magnitude, the
estimator $\hat{\range}_p^{(L)}$ over independent samples.  The gap
widens for more aggressive sampling (higher $T$).

On the IP flows and WWW logs data, there are
significant differences on the values of keys  between instances: the $L_1$ distance is a large fraction
of the total sum of values $\sum_{h\in H}\sum_{i\in[2]} v_i(h)$.
Therefore, for the destIP and Server selections, $\hat{\range}_p^{(U)}$ outperforms $\hat{\range}_p^{(L)}$ on shared-seed samples.
On the term count data there is typically a small difference between
instances.   We can see that for the 
Surnames and OSPD8 selections,
$\hat{\range}_p^{(L)}$ outperforms $\hat{\range}_p^{(U)}$ on shared seed samples.
These trends are more pronounced for the Euclidean distance ($p=2$).
In this case, on Surnames and OSPD8, $\hat{\range}_p^{(L)}$
over independent samples outperform $\hat{\range}_p^{(U)}$ over
shared-seed samples.  We can see that we can significantly improve
accuracy by tailoring the selection of the estimator to properties of
the data.  The performance of the \U\ estimator, however, can
significantly diverge for very similar instances whereas  the competitive \L\
estimator is guaranteed not to be too far off.  Therefore,  when
there is no prior knowledge on the difference,
we suggest using the \L\ estimator.

The datasets also differ in the symmetry of change.
The change is more symmetric in the IP flows and WWW logs data
$L_{p+} \approx L_{p-}$ whereas there is a general growth trend
$L_{p+} \gg L_{p-}$ in the term count data.  Estimator performance on 
one-sided distances (not shown) is similar to the corresponding
distance estimators.

\ignore{
\begin{figure*}[htbp]
\centerline{
\begin{tabular}{cc}
\ifpdf
\includegraphics[width=0.45\textwidth]{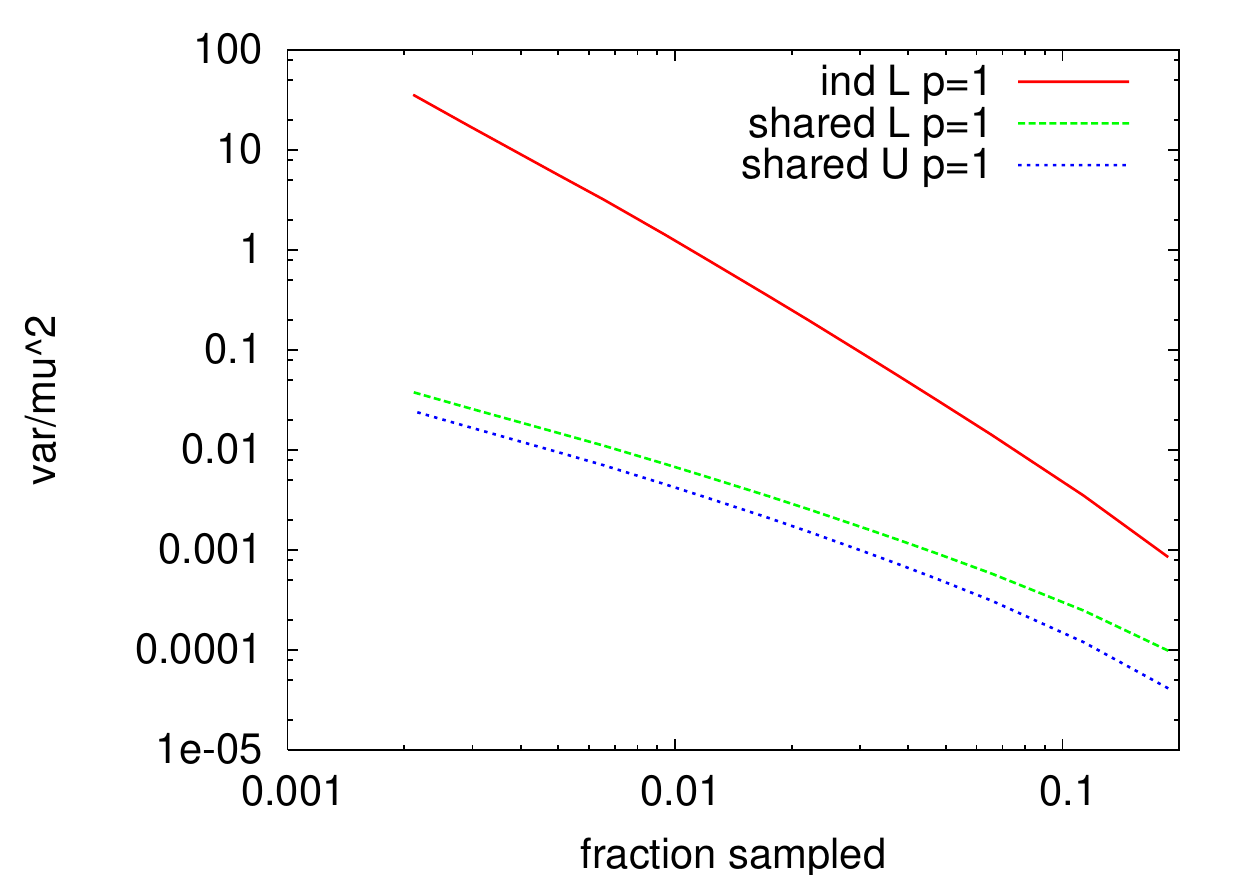}
& \includegraphics[width=0.45\textwidth]{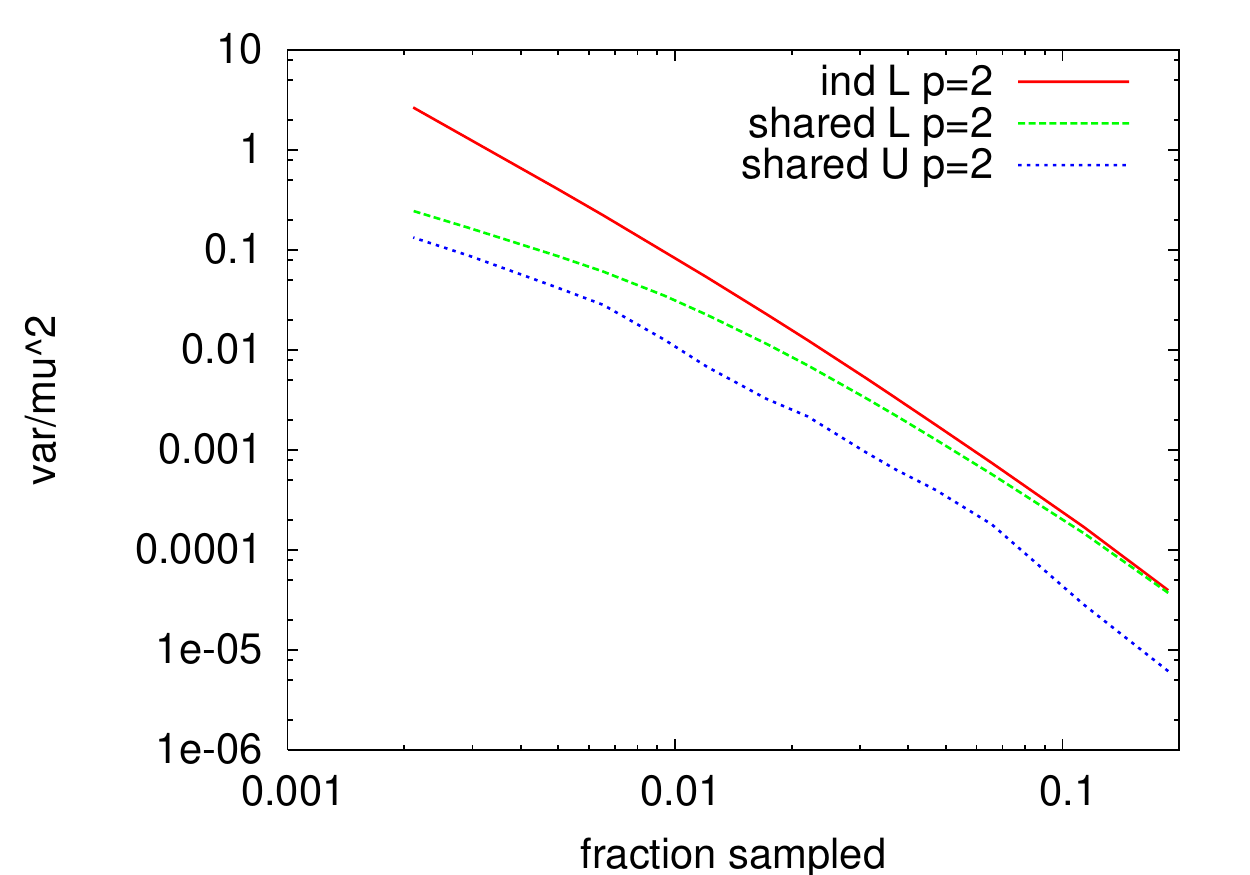} \\
\includegraphics[width=0.45\textwidth]{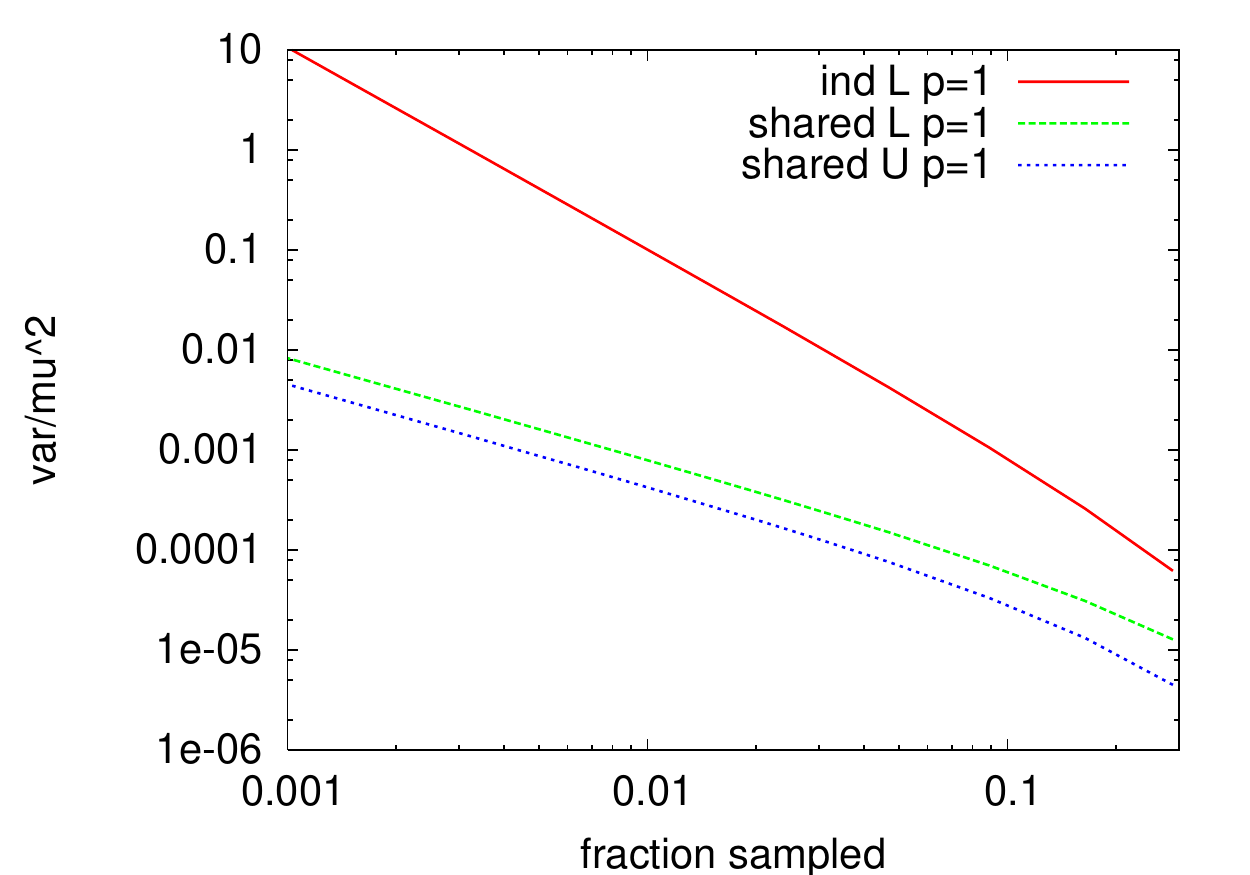}
& \includegraphics[width=0.45\textwidth]{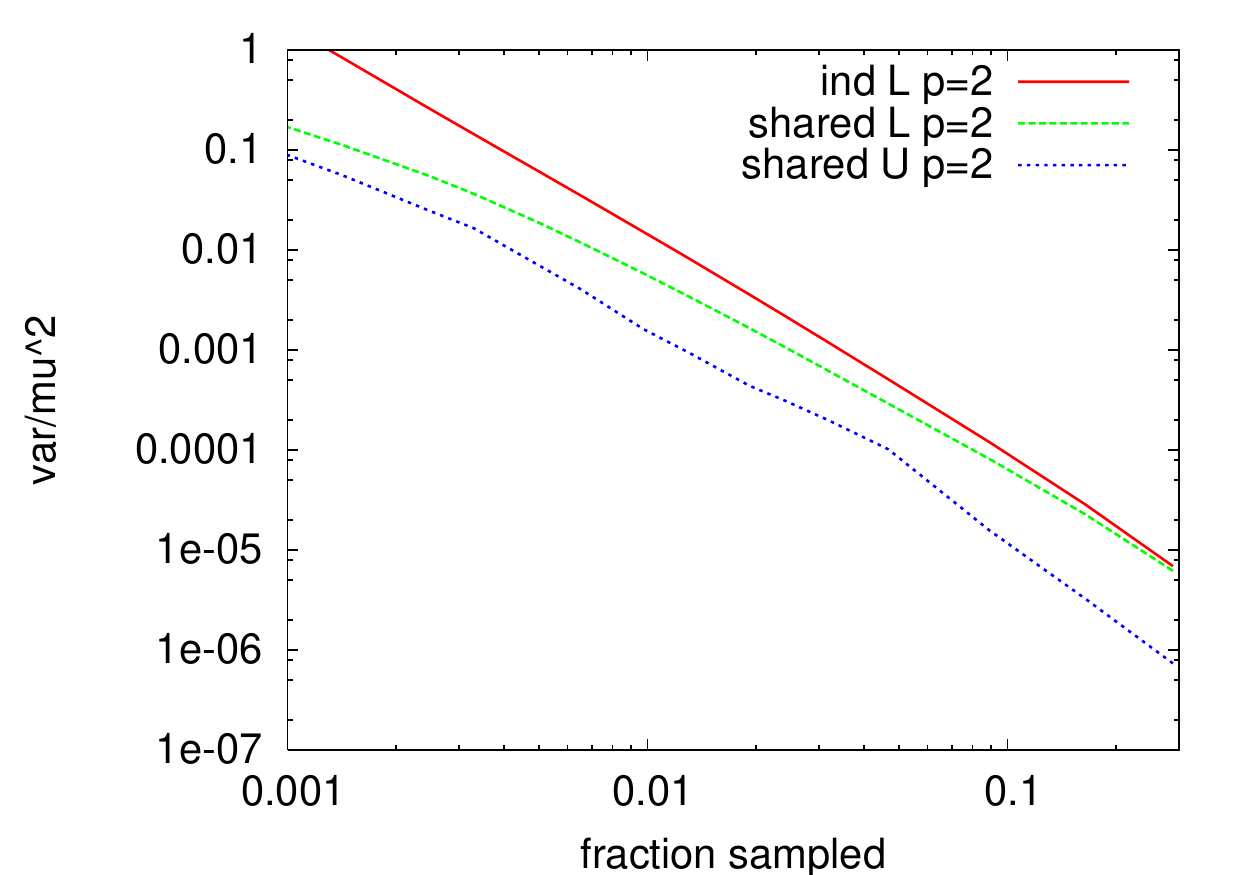} \\
\includegraphics[width=0.45\textwidth]{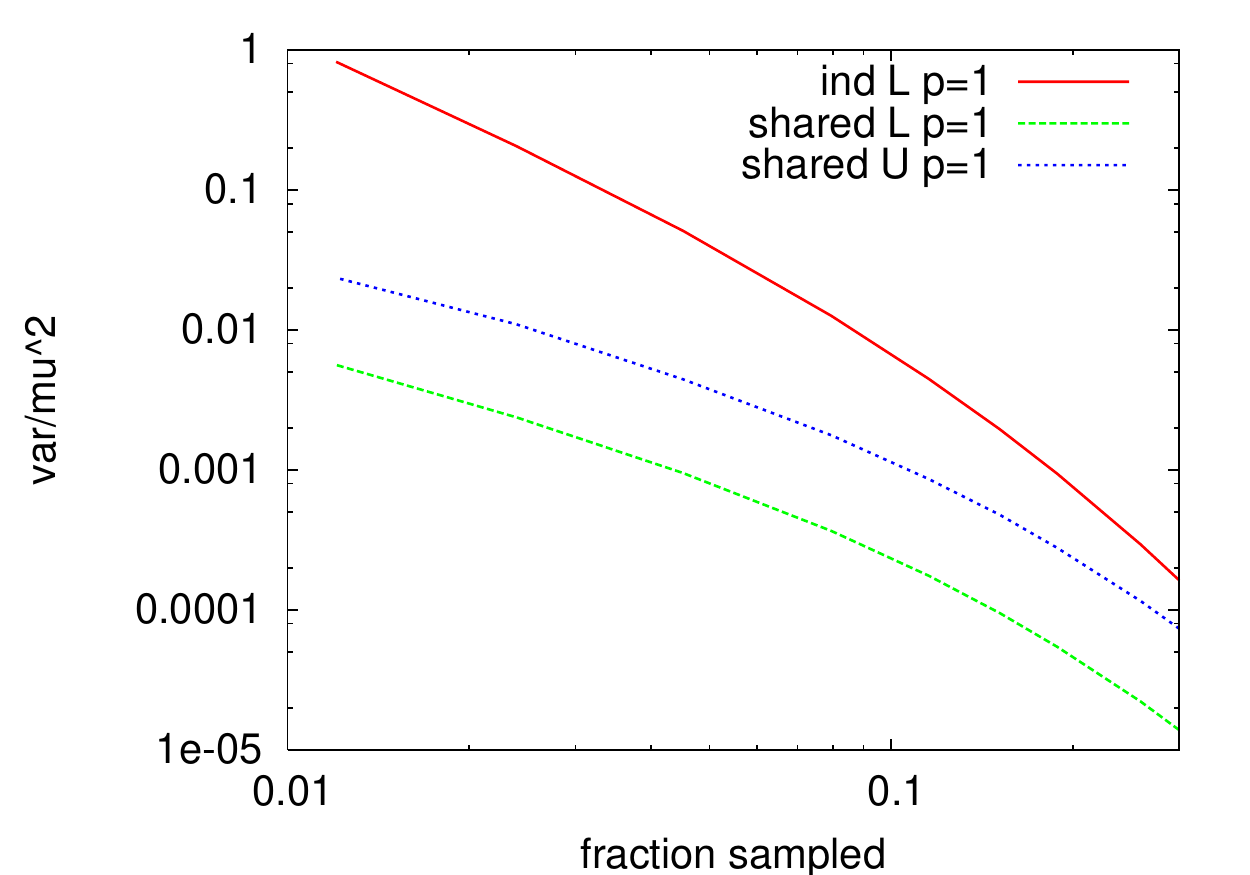}
& \includegraphics[width=0.45\textwidth]{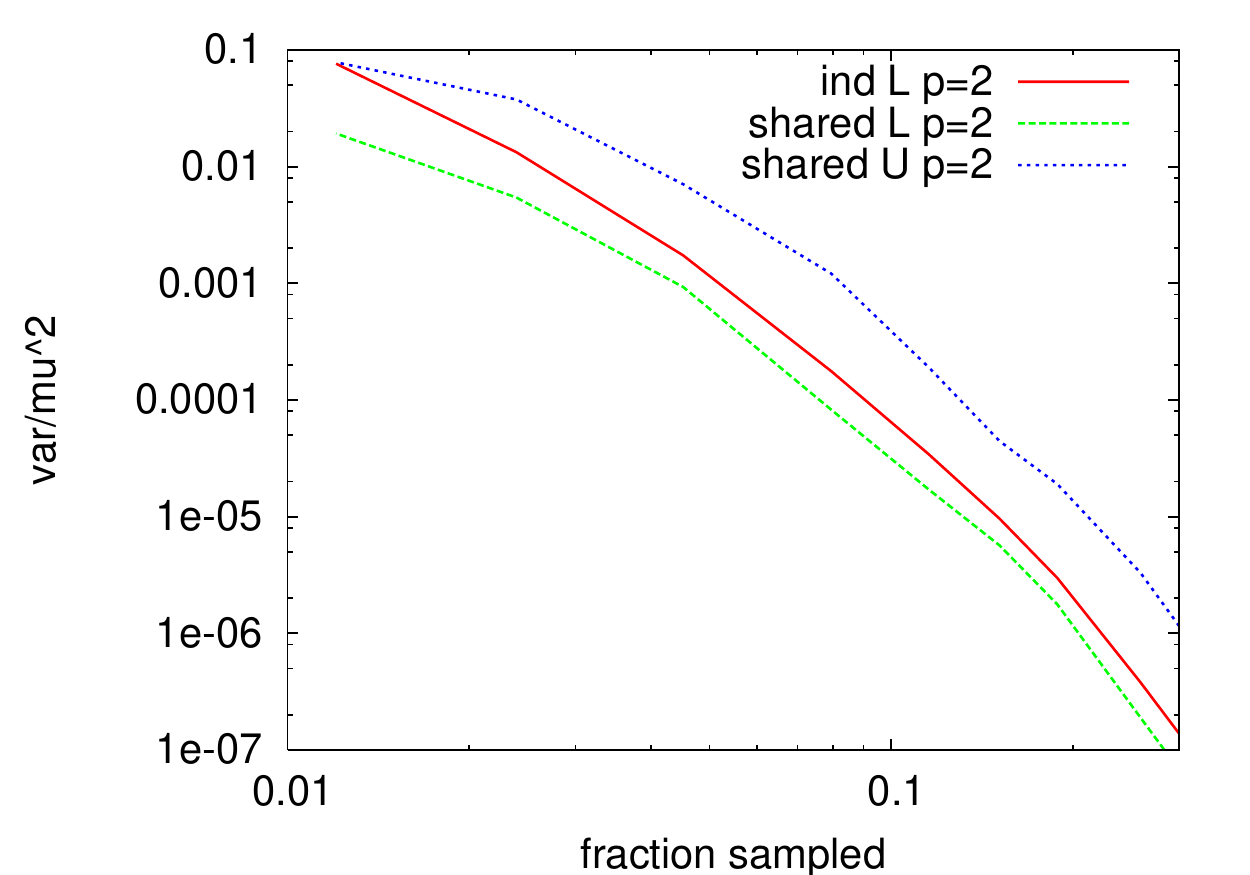} \\
\includegraphics[width=0.45\textwidth]{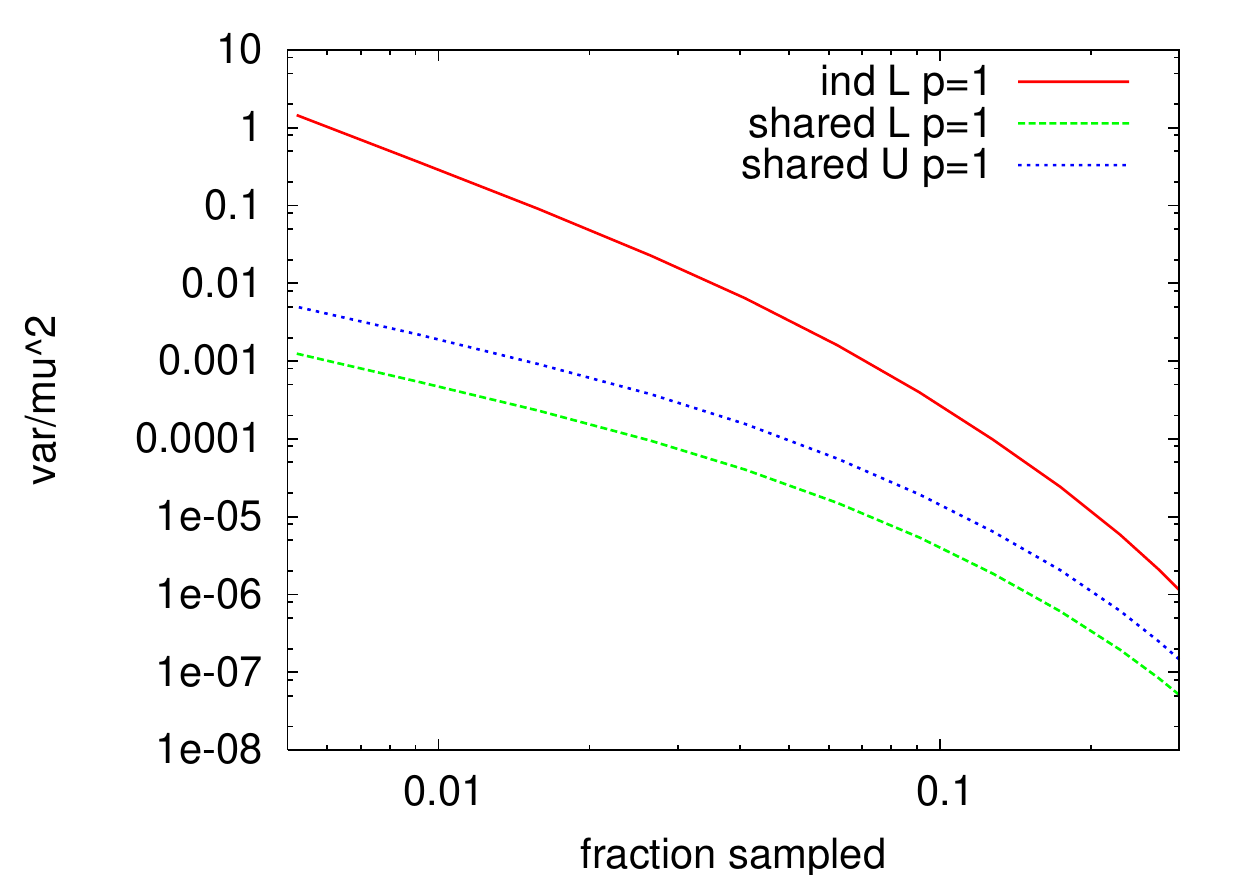} &
\includegraphics[width=0.45\textwidth]{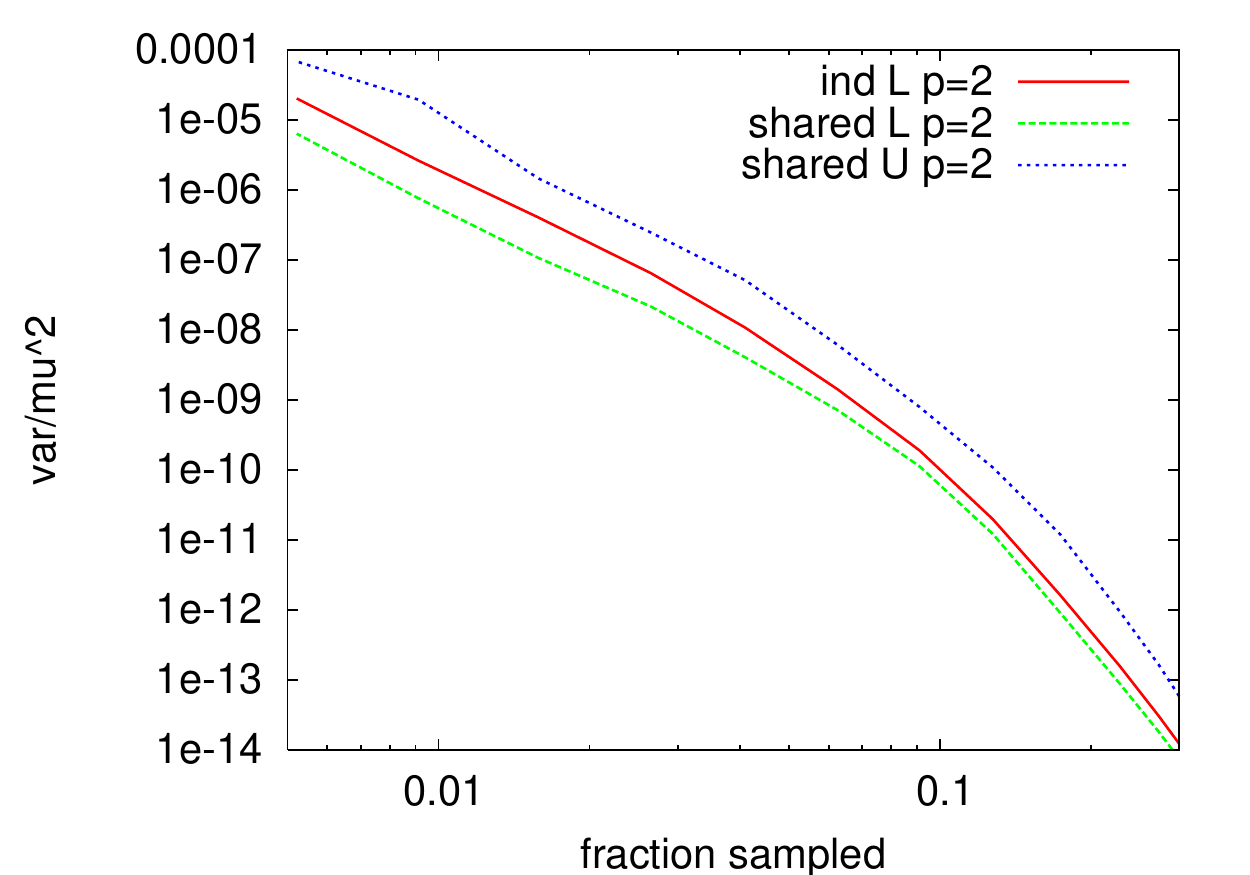} \\
\else
\epsfig{figure=dest_fcount_d1.eps,width=0.45\textwidth} & \epsfig{figure=dest_fcount_d2.eps,width=0.45\textwidth}  \\
\epsfig{figure=apache_p12_d1.eps,width=0.45\textwidth} &
\epsfig{figure=apache_p12_d2.eps,width=0.45\textwidth} \\
\epsfig{figure=surnames_2007_2008_d1.eps,width=0.45\textwidth} &
\epsfig{figure=surnames_2007_2008_d2.eps,width=0.45\textwidth} \\
\epsfig{figure=ospd_2007_2008_d1.eps,width=0.45\textwidth} &
\epsfig{figure=ospd_2007_2008_d2.eps,width=0.45\textwidth} \\
\fi
$p=1$ & $p=2$
\end{tabular}
}
\caption{Datasets top to bottom: destIP, Server, Surnames, OSPD8.\label{datasets_d12:fig}}
\end{figure*}
}

\begin{figure*}[htbp]
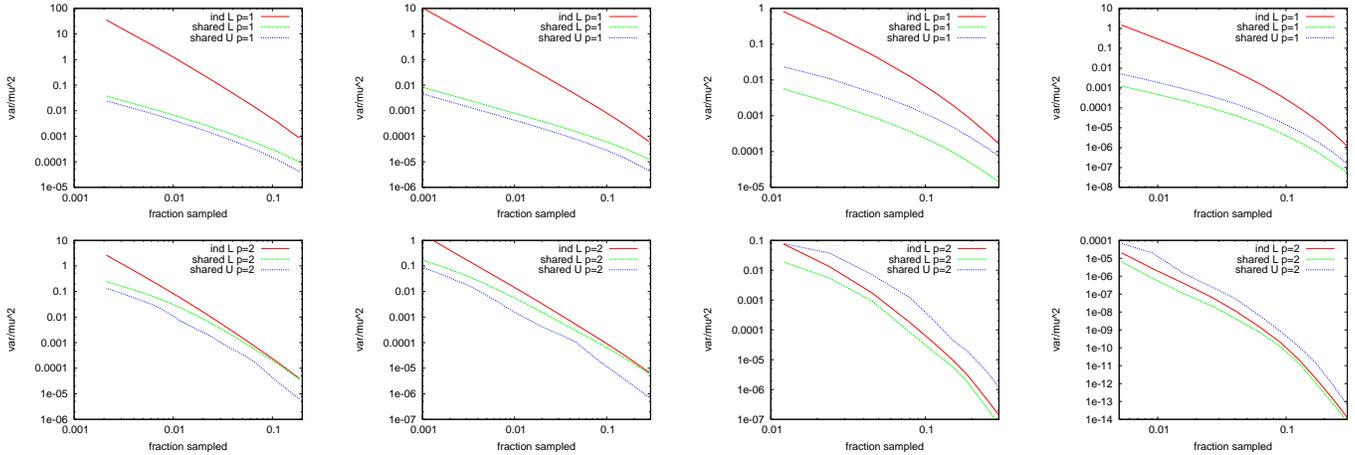

\centerline{
\begin{tabular}{cccc}
\ifpdf
\includegraphics[width=0.24\textwidth]{dest_fcount_d1} &
\includegraphics[width=0.24\textwidth]{apache_p12_d1} &
\includegraphics[width=0.24\textwidth]{surnames_2007_2008_d1} &
\includegraphics[width=0.24\textwidth]{ospd_2007_2008_d1} \\
 \includegraphics[width=0.24\textwidth]{dest_fcount_d2} &
 \includegraphics[width=0.24\textwidth]{apache_p12_d2}  &
\includegraphics[width=0.24\textwidth]{surnames_2007_2008_d2} &
\includegraphics[width=0.24\textwidth]{ospd_2007_2008_d2} \\
\else
\epsfig{figure=dest_fcount_d1.eps,width=0.24\textwidth} & 
\epsfig{figure=apache_p12_d1.eps,width=0.24\textwidth} &
\epsfig{figure=surnames_2007_2008_d1.eps,width=0.24\textwidth} & 
\epsfig{figure=ospd_2007_2008_d1.eps,width=0.24\textwidth} \\
\epsfig{figure=dest_fcount_d2.eps,width=0.24\textwidth}  &
\epsfig{figure=apache_p12_d2.eps,width=0.24\textwidth}  &
\epsfig{figure=surnames_2007_2008_d2.eps,width=0.24\textwidth} &
\epsfig{figure=ospd_2007_2008_d2.eps,width=0.24\textwidth} \\
\fi
\end{tabular}
}
\caption{Queries (left to right): destIP, Server, Surnames, OSPD8.  Plot shows CV$^2$ of $L^p_p$ estimate for fraction of sampled items from the query support. Top shows $p=1$ ($L_1$) bottom shows $p=2$ ($L_2^2$). \label{datasets_d12:fig}}
\end{figure*}

\section{Conclusion}
Distance queries are essential for monitoring, planning, and anomaly and change detection.  Random sampling is an important tool for retaining the ability to query data under resource limitations.
We provide the first satisfactory solution for estimating $L_p$
distance from sampled data sets.
Our solution is comprehensive, covering common sampling schemes.  It
is supported by rigorous analysis and novel techniques.
Our estimators scale well with data size and we demonstrated  that
accurate estimates are obtained for queries with small support size.
\paragraph*{Acknowledgement}
The author 
is grateful to Haim Kaplan for many comments, helpful feedback, and
suggesting the use of the ngrams data.

\begin{algorithm}[t]
\caption{$\hat{\range}_p^{(U)}(S)$}\label{rangedUshared:alg}
\begin{algorithmic}
\If {$|S|=0$} \Return{$0$} \Comment{from hereafter $|S|>0$}
\EndIf 
\State $\mbox{\it m}\gets \max_{i\in S} v_i$ \Comment{$\mbox{\it m}=\max(\vecv)$}
\If {$|S|<r$} $\mbox{\it n} \gets 0$
\Else $\quad \mbox{\it n}\gets \min_{i\in S} v_i$\Comment{$\mbox{\it n}=\underline{\min}(S)$}
\EndIf
\If {$\mbox{\it n}\geq \tau$} \Return $(m-n)^p$\Comment{case: $\min(\vecv)\geq \tau$}
\EndIf
\If {$p\leq 1$}\Comment{case: $\min(\vecv)\leq \tau$ and $p\leq
  1$}
\If {n=0} \Return {$m^p\frac{\tau}{\min\{m,\tau\}}$}
\Else $\quad$ \Return{$\frac{\tau}{n}\bigg(
  (m-n)^p-\frac{\min\{m,\tau\}-n}{\min\{m,\tau\}} m^p\bigg)$}
\EndIf
\EndIf
\If {$\mbox{\it m}\leq \tau$}\Comment{case: $\max(\vecv)\leq
  \tau$, $p>1$}
\If {$\zeta\tau > \mbox{\it n}$} \Return $p\tau(\mbox{\it m}-\zeta\tau)^{p-1}$
\Else $\quad$ \Return $0$
\EndIf
\EndIf
\Statex \Comment{case:  $\mbox{\it n} < \tau < \max(\vecv)$ and $p>1$}
\State $\eta_0\gets\frac{p\tau-\mbox{\it m}}{(p-1)\tau}$
\If {$\eta_0\in (0,1)$} \Comment{subcase: $\eta_0\in (0,1)$ }
\If {$\zeta \geq \max\{\eta_0, \mbox{\it n}/\tau$\}} \Return $\frac{(\mbox{\it m}-\eta_0\tau)^p}{1-\eta_0}$
\EndIf
\If {$\mbox{\it n}/\tau < \zeta < \eta_0$} \Return $p\tau(\mbox{\it m}-\zeta\tau)^{p-1}$
\EndIf
\If {$\zeta \leq \mbox{\it n}/\tau \leq \eta_0$} \Return $0$
\EndIf
\If {$\zeta \leq \mbox{\it n}/\tau \geq \eta_0 $}
\State \Return $\frac{\tau (m-n)^p}{\mbox{\it n}}-\frac{(\tau-\mbox{\it n})(\mbox{\it m}-\eta_0\tau)^p}{\mbox{\it n}(1-\eta_0)}$
\EndIf
\Else \Comment{subcase: $\eta_0\not\in (0,1)$}
\If {$\zeta\tau > \mbox{\it n}$} \Return $\mbox{\it m}^p$
\Else $\quad$
\Return $\frac{\tau}{\mbox{\it n}}(m-n)^p-\mbox{\it m}^p\bigg(\frac{\tau}{\mbox{\it n}}-1\bigg)$ 
\EndIf
\EndIf
\end{algorithmic}
\end{algorithm}

\begin{table} [h]
{\scriptsize
\begin{equation*} 
\begin{array}{ll}
 & \hat{\range}^{(L)} \\
\hline
|S|=0  \, & 0 \\
|S|\geq 1  \, & \max\{\max(\vecv)-\tau,0\}-\max\{\min(\vecv)-\tau,0\}
+\tau\ln\frac{\min\{\max(\vecv),\tau\}}{\min\{v_{\min},\tau\}}
\end{array}
\end{equation*}
\begin{equation*} 
\begin{array}{ll}
\mbox{Condition} & \var_{\mathcal{S}_{\vecv}}[\hat{\range}^{(L)}] \\
\hline
\min(\vecv) \geq \tau & 0 \\
\max(\vecv) \leq \tau,\ \min(\vecv)=0 & 2 \range(\vecv) \tau- \range(\vecv)^2\\
\max(\vecv) \leq \tau,\ \min(\vecv)>0 & 2 \range(\vecv) \tau- \range(\vecv)^2-2\tau \min(\vecv) \ln(\frac{\max(\vecv)}{\min(\vecv)})\\
0<\min(\vecv)\leq \tau \leq \max(\vecv) & (\tau)^2-\min(\vecv)^2-2\tau\min(\vecv)\ln(\frac{\tau}{\min(\vecv)})\\
0=\min(\vecv),\ \tau \leq \max(\vecv) & (\tau)^2-\min(\vecv)^2
\end{array}
\end{equation*}
}
\caption{$\hat{\range}^{(L)}$ and variance for shared-seed sampling.\label{rangeest:tab}}
\end{table}


\begin{table} [h]
{\scriptsize
\begin{equation*}
 \begin{array}{ll}
  & {\hat{\range}_2}^{(L)} \\
 \hline
 |S|=0 \,  & 0 \\
 |S|\geq 1\,  & \max\{\max(\vecv),\tau\}^2-\max\{\min(\vecv),\tau\}^2 \\
 & -2\max\{\min(\vecv),\tau\}(\max(\vecv)-v_{\min}) \\
 & +2\tau\max(\vecv)\ln\frac{\min\{\max(\vecv),\tau\}}{\min\{v_{\min},\tau\}}
 \end{array}
 \end{equation*}
}
{\scriptsize
\begin{equation*}
\begin{array}{ll}
\mbox{Condition} & \var_{\mathcal{S}_{\vecv}}[\hat{\range}_2^{(L)}] \\
\hline
\min(\vecv) \geq \tau & 0 \\
\hline
\max(\vecv) \leq \tau & -4\tau\max(\vecv)\min(\vecv) \ln(\frac{\max(\vecv)}{\min(\vecv)})(2\max(\vecv)-\min(\vecv))\\
& -(\max(\vecv)-\min(\vecv))^4\\
& +\frac{2\tau}{3}(5{\max(\vecv)}^3+4{\min(\vecv)}^3-9\max(\vecv)\min(\vecv)^2)\\
\hline
\min(\vecv) \leq \tau & 4\max(\vecv) \min(\vecv) \tau(\min(\vecv) -2\max(\vecv) )\ln\frac{\tau}{\min(\vecv)}\\
\quad\wedge &+4\max(\vecv) \min(\vecv) (\tau)^2+\frac{(\tau)^4}{3}+\frac{8{\min(\vecv)}^3\tau}{3}\\
\max(\vecv) \geq \tau & -6\max(\vecv) \min(\vecv)^2\tau-4{\max(\vecv)}^2{\min(\vecv)}^2\\
& -{\min(\vecv)}^4+4\max(\vecv) {\min(\vecv)}^3+4{\max(\vecv)}^2(\tau)^2\\
& -2\max(\vecv)(\tau)^3 \end{array}
 \end{equation*}
}  
\caption{$\hat{\range}_2^{(L)}$ and variance for shared-seed sampling\label{range2est:tab}}
\end{table}

\small{
  \bibliographystyle{plain}
  \bibliography{cycle,replace,p2p,data_structures,varopt}
}


\onlyinproc{\end{document}}


\appendix

\begin{table}
{\scriptsize
\begin{tabular}{l|l}
\hline
condition & $\hat{\range}^{(U)}$ \\
\hline
$|S|=0$ & $0$ \\
$1\leq |S|\leq r-1$ & $\max\{\tau,\max(\vecv)\}$ \\
$|S|=r$ & $\max\{\max(\vecv),\tau\}-\max\{\min(\vecv),\tau\}$
\end{tabular}
} 
 $\quad$\\
\makebox{
{\scriptsize
\begin{tabular}{l|l}
\hline
condition on $\vecv$ & $\var[\hat{\range}^{(U)}\, |\, \vecv]$ \\
\hline
$\min(\vecv)\geq \tau$ & $0$ \\
$\max(\vecv)\leq \tau$ & $\range(\vecv)(\tau-\range(\vecv))$ \\
$\min(\vecv) < \tau < \max(\vecv)$ & $\min(\vecv)(\tau-\min(\vecv))$\end{tabular}
}}
\caption{$\hat{\range}^{(U)}$ and variance for share-seed sampling.\label{rg1Uest:tab}}
\end{table}

\begin{table*}
\begin{tabular}{l|l}
\hline
condition on $S(\zeta,\vecv)$ & $\hat{\range}_2^{(U)}(S)$ \\
\hline
$\frac{\max(\vecv)}{\tau}\geq 2$, $\zeta > \frac{\min(\vecv)}{\tau}$ & $\max(\vecv)^2$\\
$\frac{\max(\vecv)}{\tau}\geq 2$, $\zeta \leq \frac{\min(\vecv)}{\tau}$ & $\max(\vecv)^2-2\tau\max(\vecv)+\min(\vecv)\tau$\\
$\frac{\max(\vecv)}{\tau}\leq 1$, $\zeta\in (\frac{\min(\vecv)}{\tau},\frac{\max(\vecv)}{\tau}]$ & $2\tau(\max(\vecv)-\zeta\tau)$\\
$\frac{\max(\vecv)}{\tau}\leq 1$, $\zeta\leq \frac{\min(\vecv)}{\tau}$ & $0$ \\
$\frac{\max(\vecv)}{\tau}\leq 1$, $\zeta\geq \frac{\max(\vecv)}{\tau}$ & $0$ \\
$\frac{\max(\vecv)}{\tau}\in [1,2]$, $\zeta > 2-\frac{\max(\vecv)}{\tau}$, $\zeta>\frac{\min(\vecv)}{\tau}$ & $4\tau(\max(\vecv)-\tau)$ \\
$\frac{\max(\vecv)}{\tau}\in [1,2]$, $\zeta < 2-\frac{\max(\vecv)}{\tau}$, $\zeta>\frac{\min(\vecv)}{\tau}$ & $2\tau(\max(\vecv)-\zeta\tau)$ \\ %
$\frac{\max(\vecv)}{\tau}\in [1,2]$, $\zeta < 2-\frac{\max(\vecv)}{\tau}$, $\zeta<\frac{\min(\vecv)}{\tau}$ & $0$\\
$\frac{\max(\vecv)}{\tau}\in [1,2]$,  $\zeta \leq \frac{\min(\vecv)}{\tau} > 2-\frac{\max(\vecv)}{\tau}$  & $\frac{\tau}{\min(\vecv)}\range_2(\vecv)-4\tau(\max(\vecv)-\tau)(\frac{\tau}{\min(\vecv)}-1)$
\end{tabular}

$\quad$\\

\begin{tabular}{l|l}
\hline
condition on $\vecv$ & $\var_{\mathcal{S}_{\vecv}}[\hat{\range}_2^{(U)}]$ \\
\hline
$\min(\vecv)\geq \tau$ & $0$ \\
$\max(\vecv)\leq \tau$ & $\range_3(\vecv)(\frac{4}{3}\tau-\range(\vecv))$\\
$\frac{\max(\vecv)}{\tau}\in [1,2]$, $\frac{\min(\vecv)}{\tau}\geq 2-\frac{\max(\vecv)}{\tau}$ & $\frac{2(\tau-\min(\vecv))}{\tau}\bigg(\range_2(\vecv))-4\tau(\max(\vecv)-\tau)\bigg)^2$\\
\hline
$\frac{\max(\vecv)}{\tau}\in [1,2]$, $\frac{\min(\vecv)}{\tau}< 2-\frac{\max(\vecv)}{\tau}$  & $\frac{\max(\vecv)-\tau}{\tau}\bigg(4\tau(\max(\vecv)-\tau)-\range_2(\vecv)\bigg)^2 + \frac{\min(\vecv)}{\tau}\range_4(\vecv)$\\
& $ + \frac{(\range_2(\vecv)+4(\tau)^2-4\max(\vecv)\tau)^3-\range_3(\vecv)(\range(\vecv)-2\tau)^3}{6(\tau)^2}$\\
\hline
$\frac{\max(\vecv)}{\tau} \geq 2$ & $(2\max(\vecv)-\min(\vecv))^2\min(\vecv)(\tau-\min(\vecv))$
\end{tabular}
\caption{Estimator $\hat{\range}_2^{(U)}$ and its variance for
  shared-seed sampling. \label{rg2Uest:tab}}
\end{table*}

\section{Derivation of $\hat{\range}_p^{(U)}$} \label{deriveRGU}

If $\zeta\tau> \max(\vecv)$ then $\underline{\range}_{p}(\zeta,\vecv)=0$ and $\hat{\range}_p^{(U)}(\zeta,\vecv)=0$.
Otherwise, when $\min(\vecv)< \zeta\tau\leq  \max(\vecv)$,
noting that the supremum is obtained by a vector $\vecv'$ with
maximum entry $\max(\vecv)$ and minimum entry $0$, 
\ignore{
we obtain
\begin{eqnarray*}
\lambda_0 & = & p\tau(\max(\vecv)-\zeta\tau)^{p-1}\\
\lambda &=& \inf_{0\leq \eta < \zeta}\frac{(\max(\vecv)-\eta\tau)^p-\int_{\zeta}^{\min\{1,\frac{\max(\vecv)}{\tau}\}} \hat{\range_p}^{(U)}(u,\vecv) d u }{\zeta-\eta}
\end{eqnarray*}
$\lambda_0= -\frac{\partial \underline{\range}_{p}(\zeta,\vecv)}{\partial \zeta}$. For $\lambda$, note that
the supremum is obtained by a vector $\vecv'$ with
maximum entry $\max(\vecv)$ and minimum entry $0$.
For $\eta\rightarrow \zeta^-$, $\underline{f}(\eta,\vecv')=\underline{f}(\eta,\vecv)=(\max(\vecv)-\eta\tau)^p$, for all $\vecv\in V(S)$ and
$\lambda \geq \lambda_0$.

If $\int_{\zeta}^1 \hat{\range}_p^{(U)}(u,\vecv)du=\underline{f}(\zeta,\vecv)$, then
$\hat{\range}_p^{(U)}(\zeta,\vecv)= \lambda_0$ else
}

\begin{eqnarray}
\lefteqn{\hat{\range}_p^{(U)}(\zeta,\vecv)\, =} \label{Urangedbetween}\\
& = & \inf_{0\leq \eta < \zeta}\frac{(\max(\vecv)-\eta\tau)^p-\int_{\zeta}^{\min\{1,\frac{\max(\vecv)}{\tau}\}} \hat{\range_p}^{(U)}(u,\vecv) d u }{\zeta-\eta}\nonumber
\end{eqnarray}
If $\zeta\tau\leq  \min(\vecv)$, 
\begin{eqnarray}
\hat{\range}_p^{(U)}(\zeta,\vecv) &=& \frac{\range_p(\vecv)-\int_{\zeta}^{\min\{1,\frac{\max(\vecv)}{\tau}\}} \hat{\range_p}^{(U)}(u,\vecv) d u }{\zeta}\nonumber\\
&=&\frac{\range_p(\vecv)-\int_{\min\{1,\frac{\min(\vecv)}{\tau}\}}^{\min\{1,\frac{\max(\vecv)}{\tau}\}} \hat{\range_p}^{(U)}(u,\vecv) d u }{\min\{1,\min(\vecv)/\tau\}}\label{Urangedunder}
\end{eqnarray}

If $\min(\vecv) \geq \tau$, $|S|=r$, $\hat{\range}_p^{(U)}=\hat{\range}_p^{(L)} \equiv \range_p(\vecv)$.
If $\max(\vecv)\leq \tau$,
$\int_\zeta^1 \hat{\range}_p(u,\vecv)du=\underline{\range}_p(\zeta,\vecv)$ and the infimum is the derivative of the lower bound function, and thus, $\hat{\range}_p(\zeta,\vecv)=p\tau(\max(\vecv)-\zeta\tau)^{p-1}$.

\begin{equation}\label{rangedestU:eq}
\begin{array}{ll}
|S| & \hat{\range}_p^{(U)} \\
\hline
0, r \, :\, & 0 \\
1\ldots r-1\, :\, & p \tau(\max(\vecv)-\zeta\tau)^{p-1}
\end{array}
\end{equation}

We now consider the case
$\min(\vecv) \leq \tau \leq \max(\vecv)$, solving \eqref{Urangedbetween}
for $\zeta > \min(\vecv)/\tau$.
For $\zeta=1$ we obtain the equation
\begin{equation*}
\hat{\range}_p^{(U)}(1,\vecv)
 =  \inf_{0\leq \eta < \zeta}\frac{(\max(\vecv)-\eta\tau)^p}{1-\eta}\ .
\end{equation*}

 When $p=1$, the derivative is positive and the infimum is
$\max(\vecv)$.  We obtain that
$\hat{\range}^{(U)}(\zeta,\vecv)=\max(\vecv)$
 for $\zeta \geq \min(\vecv)/\tau$. Using \eqref{Urangedunder},
$\hat{\range}^{(U)}(\zeta,\vecv)= \max(\vecv)-\tau$ when $\zeta\leq \frac{\min(\vecv)}{\tau}$.

For $p\not=1$, we need to find the value where
$h(\eta)=\frac{(\max(\vecv)-\eta\tau)^p}{1-\eta}$ is minimized.
The derivative is
$$\frac{\partial h(\eta)}{\partial \eta}=\frac{(\max(\vecv)-\eta\tau)^{p-1}}{1-\eta}\bigg(-\tau p+\frac{\max(\vecv)-\eta\tau}{1-\eta}\bigg)\ .$$
The derivative is $0$ at
$$\eta_0= \frac{p\tau-\max(\vecv)}{\tau(p-1)}\ .$$
If $\eta_0$ is outside $(0,1)$, the infimum is obtained at $\eta=0$ and
the estimate is
$\hat{\range}^{(U)}(\zeta,\vecv)=\max(\vecv)^p$
for $\zeta\geq \min(\vecv)/\tau$ and, using \eqref{Urangedunder},
$$\hat{\range}^{(U)}(\zeta,\vecv)=\frac{\tau}{\min(\vecv)}\range_p(\vecv)-\max(\vecv)^p(\frac{\tau}{\min(\vecv)}-1)$$
for $\zeta< \min(\vecv)/\tau$.

 Otherwise, if $\eta_0\in (0,1)$, the infimum is achieved at $\eta_0$.  Using
\eqref{Urangedbetween}, the estimate is
\begin{eqnarray*}
\hat{\range}_p(\zeta,\vecv) & = & \frac{(\max(\vecv)-\eta_0\tau)^p}{1-\eta_0}\\
& = &\frac{\tau(p-1)(\max(\vecv)\frac{p-2}{p-1}-\tau\frac{p}{p-1})^p}{\max(\vecv)-\tau}
\end{eqnarray*}
 for $\zeta\in [\max\{\eta_0,\frac{\min(\vecv)}{\tau}\},1]$ and
$\hat{\range}_p(\zeta,\vecv)=p\tau(\max(\vecv)-\zeta\tau)^{p-1}$ for $\zeta\in (\frac{\min(\vecv)}{\tau},\eta_0)$.
Using \eqref{Urangedunder}, when $\zeta \leq \frac{\min(\vecv)}{\tau}$,
then $\hat{\range}_p(\zeta,\vecv)=0$ when $\frac{\min(\vecv)}{\tau}< \eta_0$ and
$$\hat{\range}_p(\zeta,\vecv)=\frac{\range_p(\vecv)-(\frac{\min(\vecv)}{\tau}-\eta_0)\frac{(\max(\vecv)-\eta_0\tau)^p}{1-\eta_0}}{1-\frac{\min(\vecv)}{\tau}}$$ when $\frac{\min(\vecv)}{\tau}\geq  \eta_0$.

\ignore{
 The estimator for $\max(\vecv)/\tau \geq u \geq \min(\vecv)/\tau$
is
$$\frac{(\max(\vecv)-u\tau)^p}{u} - \int_{u}^{\min\{1,\frac{\max(\vecv)}{\tau}\}} \frac{(\max(\vecv)-x\tau)^p}{x^2} dx$$
The estimator for $u \leq \min(\vecv)/\tau$
is
$$(\max(\vecv)-\min(\vecv))^p\max\{1,\frac{\tau}{\min(\vecv)}\} - \int_{\min\{1,\frac{\min(\vecv)}{\tau}\}}^{\min\{1,\frac{\max(\vecv)}{\tau}\}} \frac{(\max(\vecv)-x\tau)^p}{x^2} dx$$
}



\section{Variance of {$\hat{\range}^{(U)}$} and {$\hat{\range}_2^{(U)}$}}\label{range12Ucalc:sec}

The estimators $\hat{\range}^{(U)}$ and $\hat{\range}_2^{(U)}$, provided in
Tables~\ref{rg1Uest:tab} and ~\ref{rg2Uest:tab}, are
obtained by substituting $p=1$ and $p=2$ respectively
in Algorithm~\ref{rangedUshared:alg}.
We calculate the variance of these estimators.

\noindent
{\bf Variance of $\hat{\range}^{(U)}$:}
When $\max(\vecv)\leq \tau$, we have
$\hat{\range}^{(U)}=\tau$ for
$\zeta\in (\frac{\min(\vecv)}{\tau},\frac{\max(\vecv)}{\tau}]$ and
$\hat{\range}^{(U)}=0$ otherwise. Hence,
\begin{eqnarray*}
\lefteqn{\var[\hat{\range}^{(U)}] }\\
& = & (\range(\vecv)^2(1-\range(\vecv)/\tau) + (\tau-\range(\vecv))^2 \range(\vecv)/\tau\\
& = & \range(\vecv) \tau - \range(\vecv)^2
\end{eqnarray*}

\noindent
{\bf Variance of $\hat{\range}_2^{(U)}$:}
When $\max(\vecv)>\tau$, we have
$\eta_0=2-\frac{\max\vecv}{\tau}$.  Thus $\eta_0\in (0,1) \iff \frac{\max(\vecv)}{\tau} \in (1,2)$.
We use
\begin{multline}
\int(\range_2(\vecv)-2\tau\max(\vecv)+2(\tau)^2u)^2du\\
=\frac{(\range_2(\vecv)-2\tau\max(\vecv)+2(\tau)^2u)^3}{6(\tau)^2}\ .
\end{multline}
We start with the case $\max(\vecv)\leq \tau$. \onlyinsub{Remaining
  cases are omitted due to lack of space.}
{\small
\begin{eqnarray*}
\lefteqn{\var_{\mathcal{S}_{\vecv}}[\hat{\range}_2^{(U)} ]= (1-\frac{\range(\vecv)}{\tau})\range_4(\vecv)}\\
&& + \frac{(\range_2(\vecv)-2\tau\max(\vecv)+2(\tau)^2u)^3}{6(\tau)^2}\bigg\vert^{\frac{\max(\vecv)}{\tau}}_{\frac{\min(\vecv)}{\tau}}\\
&=& (1-\frac{\range(\vecv)}{\tau})\range_4(\vecv)+ \frac{\range_6(\vecv)}{6(\tau)^2}-\frac{\range_3(\vecv)(\range(\vecv)-2\tau)^3}{6(\tau)^2} \\
&=& \range_3(\vecv)(\frac{4}{3}\tau-\range(\vecv))
\end{eqnarray*}
} 
\section{Variance of {\large $\hat{\range}^{(L)}$} and {\large $\hat{\range}_2^{(L)}$}}\label{range12Lcalc:sec}
The estimators $\hat{\range}^{(L)}$ and $\hat{\range}_2^{(L)}$,
provided in Tables~\ref{rangeest:tab} and~\ref{range2est:tab}
 are obtained using \eqref{rangedest:eq}.  We calculate their variance.

\noindent
{\bf Variance of $\hat{\range}^{(L)}$}:
 When $\max(\vecv) \leq \tau$, we have
$\hat{\range}^{(L)}=\tau \ln (\frac{\max(\vecv)}{\tau \zeta})$ when $1\leq |S|\leq r-1$ and $\hat{\range}^{(L)}=\tau \ln (\frac{\max(\vecv)}{\min(\vecv)})$ when $|S|=r$.
  The variance is
\begin{eqnarray*}
\lefteqn{\var[\hat{\range}^{(L)}|\vecv]}\\
 &=& (1-\frac{\max(\vecv)}{\tau}) \range(\vecv)^2 + \\
&& \int_{\frac{\min(\vecv)}{\tau}}^{\frac{\max(\vecv)}{\tau}} (\range(\vecv)-\tau \ln (\frac{\max(\vecv)}{\tau y}))^2 dy + \\
&& \frac{\min(\vecv)}{\tau} (\range(\vecv) - \tau \ln (\frac{\max(\vecv)}{\min(\vecv)})^2\\
&=& -2\tau \min(\vecv) \ln(\frac{\max(\vecv)}{\min(\vecv)})+2 \range(\vecv) \tau- (\range(\vecv))^2
\end{eqnarray*}



 When $\min(\vecv) \leq \tau\leq \max(\vecv)$,
$\hat{\range}^{(L)}=\max(\vecv)-\tau+\tau \ln (\frac{1}{\zeta})$ when $1\leq |S|\leq r-1$ and $\hat{\range}^{(L)}=\max(\vecv)-\tau+\tau \ln
(\frac{\tau}{\min(\vecv)})$ when $|S|=r$.
  The variance is  $\var[\hat{\range}^{(L)}|\vecv]= (\tau)^2-\min(\vecv)^2-2\tau\min(\vecv)\ln(\frac{\tau}{\min(\vecv)})$.


\ignore{
$$\hat{\range}^{(L)}=$$
$$\bigg\{\begin{array}{lc}
u > \max(\vecv)/\tau\, :\,  & 0 \\
\max(\vecv)/\tau \geq u \geq \min(\vecv)/\tau\, :\, & \tau \ln (\max(\vecv)/(u\tau)) \\
u \leq  \min(\vecv)/\tau\, :\, & \tau \ln (\max(\vecv)/\min(\vecv))
\end{array}$$

The calculation for the case $\max(\vecv)/\tau \geq u \geq \min(\vecv)/\tau$
is:
\begin{eqnarray*}
&& \frac{\max(\vecv)-u\tau}{u} -  \int_u^{\max(\vecv)/\tau} \frac{(\max(\vecv)-y\tau)}{y^2} dy \\
& = & \tau \ln (\max(\vecv)/(\tau u))
\end{eqnarray*}
}

\noindent
{\bf Variance of $\hat{\range}^{(L)}_2$:}

If $\min(\vecv)<\tau\leq \max(\vecv)$,
${\hat{\range}_2}^{(L)}=\max(\vecv)^2-(\tau)^2+ 2\tau(u\tau-\max(\vecv)+\max(\vecv)\ln\frac{1}{u}$ when $|S|\in[r-1]$ and
${\hat{\range}_2}^{(L)}=\max(\vecv)^2-(\tau)^2+ 2\tau(\min(\vecv)-\max(\vecv)+\max(\vecv)\ln\frac{\tau}{\max(\vecv)}$ when $|S|=r$.
The variance is
{\scriptsize
\begin{eqnarray*}
\lefteqn{\var[{\hat{\range}_2}^{(L)}|\vecv] =
4\max(\vecv) \min(\vecv) \tau(\min(\vecv) -2\max(\vecv) )\ln\frac{\tau}{\min(\vecv) }}\\
&&
+4\max(\vecv) \min(\vecv) (\tau)^2+\frac{(\tau)^4}{3}-6\max(\vecv) \min(\vecv)^2\tau-{\min(\vecv)}^4\\
&& -4{\max(\vecv)}^2{\min(\vecv)}^2+4\max(\vecv) {\min(\vecv)}^3+4{\max(\vecv)}^2(\tau)^2\\
&& -2\max(\vecv) (\tau)^3+\frac{8{\min(\vecv)}^3\tau}{3}
\end{eqnarray*}
}


\ignore{

                                           3
        2      3        3        2  2   8 n  T    4          2
-6 T m n  - 2 T  m + 4 n  m - 2 n  T  + ------ - n  + 4 T m n  ln(T/n)
                                          3

                             4
          2  2      2  2    T          2                2
     - 4 n  m  + 4 T  m  + ---- - 8 T m  n ln(T/n) + 4 T  n m
                            3
}

If $\max(\vecv)<\tau$, ${\hat{\range}_2}^{(L)}=2\tau(u\tau-\max(\vecv)+\max(\vecv)\ln\frac{\max(\vecv)}{u\tau})$ when $|S|\in[r-1]$ and
${\hat{\range}_2}^{(L)}=2\tau(\min(\vecv)-\max(\vecv)+\max(\vecv)\ln\frac{\max(\vecv)}{\min(\vecv)})$ when $|S|=r$.
The variance is
{\scriptsize
\begin{eqnarray*}
\lefteqn{\var[{\hat{\range}_2}^{(L)}|\vecv] = -4\tau\max(\vecv) \min(\vecv) \ln(\frac{\max(\vecv)}{\min(\vecv)})(2\max(\vecv)-\min(\vecv))+}\\
&&+\frac{2\tau}{3}(5{\max(\vecv)}^3-9\max(\vecv){\min(\vecv)}^2+4{\min(\vecv)}^3) -\range_4(\vecv)
\end{eqnarray*}
}

\end{document}